\documentclass[aip,reprint,pop]{revtex4-1}
\draft 
\usepackage{graphicx}
\usepackage{dcolumn}
\usepackage{bm}
\usepackage[utf8]{inputenc}
\usepackage[T1]{fontenc}
\usepackage{amsmath}
\usepackage{amssymb}
\usepackage{amsfonts}
\usepackage{color}
\begin{document}
\title{On collisional free-free photon absorption in warm dense
matter\thanks{File: [15\_WDM/manuscript/article\_039.tex}}
\author{J. Meyer-ter-Vehn}
\affiliation{Max-Planck-Institut für Quantenoptik,
             Hans-Kopfermann-Str. 1,
             Garching 85748, Germany}
\author{R. Ramis}
\affiliation{E.T.S.I. Aeronáutica y del Espacio,
             Universidad Politécnica de Madrid,
             P. Cardenal Cisneros 3,
             Madrid 28040, Spain}
\date{\today}
\begin{abstract}
The rate of photon absorption in warm dense matter (WDM) induced by free-free
electron-ion collisions is derived from Sommerfeld's cross-section for 
non-relativistic bremsstrahlung emission, 
making use of detailed balance relations.
Warm dense matter is treated as a metal-like state in the approximation 
of a uniform degenerate electron gas and a uniform ion background.
Total absorption rates are averaged over the electron  Fermi distribution.
A closed expression is obtained for the absorption rate, depending on 
temperature, density, and photon energy, that scales with ion charge $Z$.
It is evaluated numerically for the full parameter space, which requires 
different representations of the hypergeometric functions involved.
The results are valid for photon frequencies larger than the plasma frequency
of the medium.
They are compared with approximate formulas in various asymptotic regions.
\end{abstract}
\maketitle
%
%
\section{Introduction}
Recent achievements in generating ultra-short high-power photon and particle
pulses allow to produce small uniform volumes of warm dense matter (WDM),
a metal-like plasma state of high density and temperature in the range of the
Fermi temperature.
These plasma states play an important role in high-energy density physics
related to inertial confinement fusion \cite{Atzeni&MtV} and novel radiation
sources \cite{Norreys}. 
They are  also of fundamental interest, because they fill the gap  in the
phase diagram between  classical high-temperature plasma and solid matter.
Photon absorption by target electrons is the basic process when generating
WDM with pulses from the x-ray free electron lasers (XFEL), e.g. now available
at Stanford \cite{Stanford_XFEL} and DESY \cite{DESY_XFEL}.
The XFEL beams are used for both generating and probing WDM.
Also laser-generated beams of high laser harmonics have been used to probe
free-free absorption in solid aluminum \cite{Hollebon}; the comb of high harmonics
allows to measure absorption at different frequencies in a single shot.

The present paper is devoted to collisional absorption of soft X-rays in WDM,
which is the free-free part of the total absorption.
Though absorption by bound-free and bound-bound transitions (in case they are
admitted) often dominates the total rate, here we restrict ourselves to the
contribution of free electrons and consider, in particular, the case of fully
ionized plasmas.

The central idea of the present paper is to derive the rate of photon
absorption from the cross-section for bremsstrahlung, making use of detailed
balance relations.
The differential cross-section of bremsstrahlung emitted during Coulomb
scattering of electrons by ions  was first derived quantum-mechanically by
Sommerfeld \cite{Sommerfeld}.
It is valid not only in the limit of fast electrons, where $Ze^2/\hbar v\ll1$
and the Born approximation apply, but also covers slow electrons in
the region  $Ze^2/\hbar v \approx 1$, which is relevant for WDM and where
Coulomb scattering is strongly modified. Here $Ze$ is the ion charge,
$\hbar=h/(2\pi)$ Planck's constant, and $v$ the electron velocity. 
The rate of spontaneous photon absorption is then corrected for stimulated
emission and averaged over the Fermi distribution, describing the degenerate
electrons in WDM.
This provides the coefficient for total free-free photon absorption.
It depends on three parameters: temperature and density of electrons as well
as photon energy.

The procedure makes use of the Drude-Sommerfeld model; for a modern
presentation, see the book of Ashcroft and Mermin on solid state physics
\cite{Ashcroft&Mermin}. 
It describes electrons as a Fermi gas of independent particles that collide
with ions.
The collision frequency $\nu$ was used as a parameter and adjusted to
experiments.
In the present approach, however, we consider inverse bremsstrahlung 
collisions explicitly to describe photon absorption. 
The asumption of a uniform Fermi gas is certainly a rough approximation 
for WDM, in particular when approaching zero temperature and high densities 
above solid density. 
For these regions, we cannot expect that the present approach leads to 
quantitative agreement with experiments.
However, it allows us to obtain closed formulas and scaling relations 
covering a wide region of the parameter space. 
This is a major objective of the present paper. 
It may serve as a general reference when using more sophisticated theory.
Making use of the quantum-statistical approach to WDM, optical properties 
have been treated in a number of papers (see e.g. Ref. \onlinecite{Redmer} and 
papers quoted therein). 
Recently, density functional theory has been successfully applied to 
absorption measurements in solid aluminum at low temperatures and 
frequencies below the L-shell edge, 
where only valence electrons contribute \cite{Hollebon}. 
Also in order to account for effects at high density, the 
temperature-dependent Thomas-Fermi model has been used \cite{Ishikawa1998}.

In the present paper, we restrict ourselves to frequencies above the plasma 
frequency, $\omega>\omega_p$, where radiant collisions occur close to the 
ions at distances $\sim v/\omega$ smaller than the Debye length and 
collective plasma effects like screening of the Coulomb potential or 
ion-ion correlations play a minor role. 
Here $v$ is a typical electron velocity such as the Fermi velocity, 
$\omega_p=\sqrt{4\pi e^2n/m}$ is the plasma frequency, $n$ the electron 
number density, and $m$ the electron mass.  
Under the present assumptions, we make use of linear plasma theory.
The photon absorption coefficient $\alpha$ is then related to the 
collision frequency by
   \begin{equation}\label{alpha}
   \alpha=\frac{\nu}{cn_R}\frac{\omega_p^2}{\omega^2},
   \end{equation}
where  $n_R=(1-\omega_p^2/\omega^2)^{1/2}$ is the index of refraction
and $c$ is the velocity of light \cite{Atzeni&MtV}.
For dilute high-temperature plasma, the Spitzer collision frequency
\cite{Spitzer}
   \begin{equation}\label{Spitzer}
   \nu_{\rm{Spitzer}}=\frac{4\sqrt{2\pi}}{3}\frac{Zne^4}{\sqrt{m}(kT)^{3/2}}
   \ln\frac{2\sqrt{3}kT}{\hbar\omega_p}
   \end{equation}
is typically used to describe electron-ion collisions.
It diverges for $kT\rightarrow 0$ and does not apply to WDM.
The present approach reproduces the Spitzer formula as an upper limit for
high temperatures and $\omega\rightarrow\omega_p$.
It also naturally
describes the transition $kT\rightarrow 0$ without introducing any ad hoc
cut-offs.
The results will be presented in terms of an effective dynamic collision
frequency $\nu_{\rm{eff}}(\omega)$. 
Finite values of $\nu_{\rm{eff}}$ are not only 
due to Fermi degeneracy, but also due to a smooth transition 
from 'fast' to 'slow' electron scattering, implicit in the
fully quantum-mechanical treatment.

The paper is organized as follows: 
In Sec. II, the central formula for the effective collision frequency
$\nu_{\rm eff}(\omega)$ in warm dense matter is derived.
Sec. III is devoted to the numerical evaluation and a discussion of
$\nu_{\rm eff}$. 
Sec. IV provides formulas for various asymptotic regions.
Appendix A describes how the hypergeometric function is evaluated 
in various parameter regions. 
In Appendix B approximate formulas are derived 
for different asymptotic regions. Appendix C deals with numerical integration.

Only in the final stages of preparing this manuscript, we became aware of
parallel work in the astrophysical literature, concerned with radiative
energy transport in stellar interiors as well as interstellar matter.
Following the pioneering work of Kramers in 1923 \cite{Kramers} and Gaunt 
in 1930 \cite{Gaunt1930}, it has become common practice in astrophysics to
describe radiative Coulomb collisions by Kramers's quasi-classical  
cross-section times a quantum-mechanical correction, the so-called Gaunt
factor.
As is turns out, Sommerfeld's cross-section, on which the present work is
based, can also be represented in this way, thus providing a general
expression for the free-free Gaunt factor.
Following the seminal paper by Karzas \& Latter \cite{KL1960}, recent 
papers \cite{Sutherland1998, Hoof2014, Avillez2015} have been devoted 
to an accurate numerical determination of the free-free Gaunt factor.
They present results similar to ours.
An important difference is that thermal averages are discussed only for
Maxwell  distributions, adequate for the envisioned astrophysical applications, 
while the present focus is on degenerate plasmas which requires averages 
over Fermi distributions.
Also we discuss in detail the asymptotic limits.
We expect that a careful comparison with the astrophysical work, not yet
given here, will lead to fruitful extensions of the present work.
%
%
\section{Derivation of the collisional photon absorption rate}
\subsection{The cross-section for bremsstrahlung}
We start from Sommerfeld's quantum-mechanical result 
for the differential cross-section of spontaneous bremsstrahlung emission
\cite{Sommerfeld}. 
It describes the scattering of a non-relativistic electron 
with momentum $p$ on an ion with charge $Ze$ under emission
of a photon with frequency $\omega$, averaged over emission angles  
and polarizations of the photon. In the main part of this paper, we shall 
use the notation $p_\pm=\sqrt{p^2\pm 2m\hbar\omega}$ for the momentum of 
the scattered electron, depending on whether the photon is absorbed or emitted. 
This notation is also used for the variables $\eta=Z\alpha_fmc/p$ and 
$\eta_\pm=Z\alpha_fmc/p_\pm$, characterizing Coulomb wave functions.
With the fine-structure constant $\alpha_f=e^2/\hbar c$ and Bohr's radius
$a_B=\hbar^2/(me^2)$ the cross-section is given by
\begin{widetext}
   {
   \begin{equation}\label{Sommerfeld cross-section}
   \frac{d\sigma}{d\omega}(p\rightarrow p_-,\omega)=
   \frac{64\pi^2}{3} \frac{ Z^2\alpha_f^5 a_B^2}{\omega}
   \left(\frac{mc}{p}\right)^2\cdot
   \frac{\xi}{4}\frac{d|F(i\eta_-,i\eta;1;\xi)|^2/d\xi}
   {(1-e^{-2\pi\eta_-})(e^{2\pi\eta}-1)};
   \end{equation}
   }
\end{widetext}
here $F(i\eta_-,i\eta;1;\xi)$ denotes the complete hypergeometric function
and $\xi=-4\eta\eta_-/(\eta-\eta_-)^2$.

In passing, we mention already here that Eq.~(\ref{Sommerfeld cross-section}) 
reduces to Kramers's cross-section \cite{Kramers}
   \begin{equation}\label{Kramers}
   \bigg(\frac{d\sigma(p\rightarrow p_-,\omega)}{d\omega}\bigg)_{\rm{Kram}}=
   \frac{16\pi}{3\sqrt{3}}
   \frac{Z^2\alpha_f^5a_B^2}{\omega}\bigg(\frac{mc}{ p}\bigg)^2
   \end{equation}
in quasi-classical approximation, i.e. for small electron momenta such that
$\eta\gg1$ and $(\eta_--\eta)\gg 1$.
Small electron momenta in the range of the Fermi
momentum play a dominant role in WDM; 
asymptotic approximations of Eq.~(\ref{Sommerfeld cross-section}) 
will be discussed in detail further below.
We notice that Eq.~(\ref{Sommerfeld cross-section}) can be written as
   \begin{equation}
   \frac{d\sigma}{d\omega}(p\rightarrow p_-,\omega)=
   \bigg(\frac{d\sigma(p\rightarrow p_-,\omega)}{d\omega}\bigg)_{\rm{Kram}}
   g_{\rm{ff}},
   \end{equation}
where
   \begin{equation}\label{Gaunt factor}
   g_{\rm{ff}}=4\pi\sqrt{3}\cdot\frac{\xi}{4}
   \frac{d|F(i\eta_-,i\eta;1;\xi)|^2/d\xi}{(1-e^{-2\pi\eta_-})(e^{2\pi\eta}-1)}
   \end{equation}
is the so-called free-free Gaunt factor, used in the astrophysical literature
to describe quantum-mechanical corrections to Kramers's quasi-classical
cross-section.
%
%
\subsection{Detailed balance between photon absorption and photon emission}

The central point of this paper is now to derive the rate of collisional
photon absorption $R_{\rm{abs}}(p,\omega\rightarrow p_+)$;
it gives the number of photons with energy $\hbar\omega$ absorbed
per unit of time by an electron with momentum $p$ that  is boosted to
momentum $p_+=\sqrt{p^2+2m\hbar\omega}$, while scattering on an ion with
charge $Ze$.
This process is inverse to bremsstrahlung emission and is therefore also
called absorption by \textit{inverse} bremsstrahlung.
The corresponding rate of bremsstrahlung emission can be calculated from
Sommerfeld's cross-section; it is given by
   \begin{equation}\label{R_em}
   R_{\rm{em}}(p_+\rightarrow p,\omega)=\frac{p_+}{m}n_i
   \frac{d\sigma(p_+\rightarrow p,\omega)}{d\omega}\Delta \omega
   \end{equation}
and determines the number of photons with frequency between $\omega$ and
$\omega+\Delta \omega$ emitted per unit of time when an electron with
momentum $p_+$ passes matter having ion density $n_i$.

The rates for absorption and emission are related by the principle of
detailed balance, which is discussed in many textbooks on statistical
mechanics (see e.g. Ref. \onlinecite{Reif}).
It was used  by Einstein in the context of optical transitions to discover
stimulated emission \cite{Einstein}, the cornerstone of laser physics.
Detailed balance relations are derived for systems in statistical
equilibrium, but they are valid under much more general conditions.
On a microscopic level, detailed balance is related to time reversal symmetry.
A brief discussion for practical use is given in Ref. \onlinecite{Atzeni&MtV}.
For the ratio of the rates for absorption and emission, one obtains
   \begin{equation}\label{detailed balance}
   \frac{R_{\rm{abs}}(p,\omega\rightarrow p_+)}{R_{\rm{em}}
   (p_+\rightarrow p, \omega)}
   =\frac{dZ(\varepsilon_+)/d\varepsilon_+}{dZ(\varepsilon)/
   d\varepsilon\cdot Z_{\rm{ph}}};
   \end{equation}
it depends only on the number of quantum states available for an electron
and/or a photon in the final states of the two processes.
For an electron of energy $\varepsilon=p^2/2m$  (and $\varepsilon_+=p_+^2/2m$,
respectively), one finds
   \begin{equation}\label{phase factor E}
   dZ(\varepsilon)/d\varepsilon =(2/h^3)V_e 4\pi p^2 dp/d\varepsilon 
   =(8\pi m V_e/h^3)p,
   \end{equation}
where $2/h^3$ is the density of quantum states available in phase space
for electrons with 2 spin directions and $V_e=1/n$ is the volume available
for one electron.
The number of quantum states $Z_{\rm{ph}}$ available for a photon in the
final state, having a frequency between $\omega$ and $\omega+\Delta\omega$, is
   \begin{equation}
   Z_{\rm{ph}}=(2/h^3)V_{ph}4\pi(\hbar\omega/c)^2(\hbar\Delta\omega/c)=
   (\omega^2\Delta\omega)/(\pi^2c^3)V_{\rm{ph}},
   \end{equation}
where the photon volume $V_{\rm{ph}}=((E_0^2/8\pi)/\hbar\omega)^{-1}$ is the
inverse of the photon number density and $E_0$ is the amplitude 
of the electric field of the light wave.
Actually, $Z_{\rm{ph}}$ is the inverse of the photon number per quantum state
   \begin{equation} \label{photons per quantum state}
   n_{\rm{ph}}=\frac{1}{Z_{\rm{ph}}}=\frac{\pi^2c^3}{\omega^2\Delta\omega}
   \cdot\frac{E_0^2/8\pi}{\hbar\omega},
   \end{equation}
a quantity needed in the following to describe stimulated emission.
From Eqs. (\ref{detailed balance} - \ref{photons per quantum state})
we obtain the important result
   \begin{equation}\label{Rabs}
   R_{\rm{abs}}(p,\omega\rightarrow p_+)=\frac{p_+}{p} n_{\rm{ph}} 
   R_{\rm{em}}(p_+\rightarrow p, \omega).
   \end{equation}
%
%
\subsection{The photon absorption density in WDM}
We now turn to warm dense matter and show how to apply the relations derived
above to a dense degenerate electron gas.
First the absorption rate has to be averaged over the Fermi distribution
function for the electrons
   \begin{equation} \label{chem potential}
   f(p)=\frac{1}{1+\exp{\big((p^2/2m-\mu)/kT\big)}},
   \end{equation}
where $kT$ is the electron temperature and the chemical potential $\mu$
is determined implicitly by electron density
   \begin{equation} \label{chem potential}
   n=\int_0^\infty (2/h^3)4\pi p^2dp f(p). 
   \end{equation}
The integrated absorption rate per volume then is
   \begin{equation}
   A_{\rm{abs}}=\int_0^\infty\frac{2}{h^3} R_{\rm{abs}}(p,\omega\rightarrow p_+)
   f(p)\big(1-f(p_+)\big)4\pi p^2dp;
   \end{equation}
here we account for the case that the final electron state with
momentum $p_+$ is partly occupied in the Fermi gas.
Let us also account for stimulated emission of photons with frequency $\omega$, 
which lowers the electron momentum from $p$ to 
$p_-=\sqrt{p^2-2m_e\hbar\omega}$; it occurs at a rate
   \begin{equation} \label{stim emission}
   R_{\rm{em}}^{\rm{stim}}(p\rightarrow p_-,\omega)
   =n_{\rm{ph}}R_{\rm{em}}(p\rightarrow p_-,\omega).
   \end{equation}
This stimulated emission reduces overall absorption.
The total absorption rate per volume is therefore obtained as
   \begin{equation}\label{total absorption rate}
   A_{\rm{abs}}^{tot}=\int_0^\infty \frac{2}{h^3}
   R_{\rm{abs}}(p,\omega\rightarrow p_+)f(p)(1-f(p_+))4\pi p^2dp$$$$
   \quad  -\int_{p_{\rm{min}}}^\infty \frac{2}{h^3}
   R_{\rm{em}}^{\rm{stim}}(p\rightarrow p_-, \omega)f(p)(1-f(p_-))4\pi p^2dp
   \end{equation}
with $p_{\rm{min}}=\sqrt{2m\hbar\omega}$.
Now making use of Eq.~(\ref{Rabs}) and Eq.~(\ref{stim emission}) and
changing integration variables appropriately, we can combine the two
integrals and find
   \begin{equation}\label{A_abstot}
   A_{\rm{abs}}^{tot}=\frac{8\pi}{h^3}n_{\rm{ph}} \int_0^\infty    
   p_+ R_{\rm{em}}(p_+\rightarrow p,\omega) \big( f(p) - f(p_+) \big) p dp.
   \end{equation}
%
%
\subsection{The effective collision frequency for warm dense matter}
Finally, we relate $A_{\rm{abs}}^{\rm{tot}}$ to the photon absorption 
coefficient $\alpha$  introduced in Eq.~(\ref{alpha}) and to an effective 
dynamic collision frequency $\nu_{\rm eff}(\omega)$.
For this, we notice that $A_{\rm{abs}}^{\rm{}tot}$  can be  expressed also as
   \begin{equation}
   A_{\rm{abs}}^{\rm{tot}}=\Phi_{\rm{ph}} \alpha=\Phi_{\rm{ph}} 
   \bigg(\frac{\nu_{\rm{eff}}}{cn_R}\bigg)\frac{\omega_p^2}{\omega^2},
   \end{equation}
where $\Phi_{\rm{ph}}=(E_0^2/(8\pi\hbar\omega))c n_R$ denotes the flux
of photons, which is given by the photon density $E_0^2/(8\pi\hbar\omega)$ 
times the group velocity $cn_R$ of the light pulse close to the scattering ion.
This allows us to define the effective collision frequency
   \begin{equation}
   \nu_{\rm{eff}}(\omega)=\frac{\omega^2}{\omega_{\rm{p}}^2}\frac{\hbar\omega}
   {E_0^2/8\pi} A_{\rm{abs}}^{\rm{tot}}.
   \end{equation}
As it turns out, $\nu_{\rm{eff}}(\omega)$ is a convenient quantity
to express the results of this paper.
Making use of  Eqs.~(\ref{R_em}), (\ref{photons per quantum state}), 
and (\ref{A_abstot}), we are led to the central result of the present paper:
   \begin{equation}\label{nu_eff}
   \nu_{\rm{eff}}(\omega)=\frac{n_ic^3}{\hbar^3\omega_p^2 m}\int_0^\infty
   p_+^2 \frac{d\sigma(p_+\rightarrow p,\omega)}{d\omega}
   \big( f(p)-f(p_+) \big) p dp
   \end{equation}
This closed expression allows us to calculate the effective dynamic collision
frequency for photon absorption in warm dense matter from the
cross-section for bremsstrahlung given in Eq.~(\ref{Sommerfeld cross-section}).
By means of Eq.~(\ref{alpha}), it provides also the collisional absorption
coefficient $\alpha$ and the free-free opacity $\kappa=\alpha/\rho$,
where $\rho$ is the mass density of the absorbing medium.

For further discussion, let us rewrite  the cross-section
Eq.~(\ref{Sommerfeld cross-section})
   \begin{equation}\label{SCC}
   \frac{d\sigma}{d\omega}(p_+\rightarrow p,\omega)
   =\frac{64\pi^2}{3}
   \frac{Z^2\alpha_f^5a_B^2}{\omega}\bigg(\frac{mc}{ p_+}\bigg)^2\cdot
   G(\hat{\varepsilon},\hat{\omega}),
   \end{equation}
in terms of the dimensionless function
   \begin{equation}\label{G}
   G(\hat{\varepsilon},\hat{\omega})=\frac{\xi}{4}\frac
   {d|F(i\eta,i\eta_+;1;\xi)|^2/d\xi}{(1-e^{-2\pi\eta})(e^{2\pi\eta_+}-1)},
   \end{equation}
where the coordinates are related to each other by
   \begin{equation}
   \hat{\varepsilon}=\frac{\varepsilon}{Z^2E_a}=\frac{p^2}{2mZ^2E_a}
   =\frac{1}{2\eta^2},
   \end{equation}
   \begin{equation}
   \hat{\omega}=\frac{\hbar\omega}{Z^2E_a},
   \end{equation}
   \begin{equation}
   \hat{\varepsilon}+\hat{\omega}=\frac{\varepsilon+\hbar\omega}{Z^2E_a}
   =\frac{p_+^2}{2mZ^2E_a}=\frac{1}{2\eta_+^2}.
   \end{equation}
In terms of function $G(\hat{\epsilon},\hat{\omega})$, the effective collision
frequency can be written in particularly transparent form:
   \begin{widetext}
   \begin{equation}\label{nu_final}
   \nu_{\rm{eff}}(\omega)= \frac{16\pi}{3}Z\nu_0
   \int_0^{\infty} G\bigg(\frac{\varepsilon}{Z^2E_a},
   \frac{\hbar\omega}{Z^2E_a}\bigg)
   \bigg[ f\bigg(\frac{\varepsilon-\mu}{kT}\bigg)
   -f\bigg(\frac{\varepsilon+\hbar\omega-\mu}{kT}\bigg)\bigg]
   \frac{d\varepsilon}{\hbar\omega},
   \end{equation}
   \end{widetext}
showing explicitly the dependence on ion charge $Z$, temperature $kT$, and
chemical potential $\mu$, which is related implicitly to the electron
density $n$.
Here the Fermi function is given by $f(x)=1/(1+e^x)$,
$\nu_0=E_a/\hbar=4.14\times 10^{16}$$s^{-1}$ is the atomic frequency unit,
and $E_a=\alpha_f^2mc^2=me^4/\hbar^2=27.2$ eV the atomic energy unit.
The key quantity is the kernel function $G(\hat{\epsilon},\hat{\omega})$,
depending on electron energy and photon energy, both normalized to $Z^2E_a$.
The kernel $G(\hat{\epsilon},\hat{\omega})$ describes radiative 
Coulomb collisions in the full non-relativistic region.
It is equivalent to the Gaunt factor defined in Eq.~(\ref{Gaunt factor}).
%
%
\section{Numerical evaluation}

The numerical evaluation of Eq.~(\ref{nu_final}) is not straightforward.
This is because $G(\hat{\varepsilon},\hat{\omega})$  involves 
the hypergeometric function.
We develop the required representations for the different parameter
regions in Appendix A, based on the Handbook for Mathematical 
Functions \cite{abramovitz,nist}. 
Asymptotic approximations are found in Refs.
\onlinecite{Sommerfeld,Landau&Lifshitz,KrainovI,KrainovII}, and corresponding
expressions are derived in Appendix B.
%
%
   \begin{figure}[t]
   \begin{center}
   \includegraphics[width=\columnwidth]{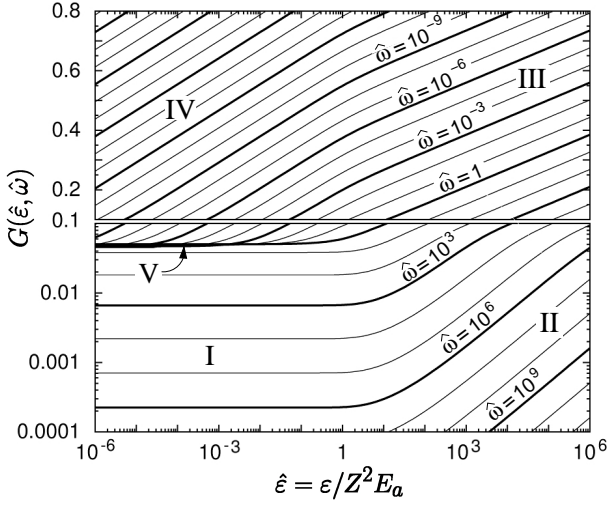}
   \end{center}
   \caption{\label{Fig1}
   Function $G(\hat\varepsilon,\hat\omega)$ plotted versus
   $\hat{\varepsilon}$ for fixed values of $\hat{\omega}$, 
   given as labels in steps of factor 10.
   Notice that the ordinate scale is logarithmic 
   for $G\le 0.1$ and linear for $G\ge 0.1$. 
   The Roman numerals identify parameter regions in which 
   $G(\hat\varepsilon,\hat\omega)$ allows for simple asymptotic formulas.}
   \end{figure}
%
%
   \begin{table}[]
   \begin{ruledtabular}
   \caption{Asymptotic expressions of $ G(\hat\varepsilon,\hat\omega)$ for the
   five regions marked in Fig.~\ref{Fig1}; $\gamma=0.577$ is the Euler number.}
   \label{table}
   \begin{center}
   \begin{tabular}{rll}
   I & $\hat\varepsilon\ll 1$, $\hat\omega\gg 1$ &
   $G(\hat\varepsilon,\hat\omega)=
   \displaystyle{\frac{1}{\pi\sqrt{2\hat\omega}}}$ \\
   &&\\
   II & $\hat\varepsilon\gg 1$, 
   $\hat\omega\gg \hat\varepsilon$ &
   $G(\hat\varepsilon,\hat\omega)=
   \displaystyle{\frac{1}{2\pi^2}\sqrt{\frac{\hat\varepsilon}{\hat\omega}}}$\\
   &&\\
   III & $\hat\varepsilon\gg 1$,
   $\hat\omega\ll \hat\varepsilon$ &
   $G(\hat\varepsilon,\hat\omega)=\displaystyle{
   \frac{1}{4\pi^2}
   \ln\frac{4\hat\varepsilon}{\hat\omega}}$\\
   &&\\
   IV  & $\hat\varepsilon\ll 1$,
   $\hat\omega\ll \hat\varepsilon^{3/2}$ &
   $G(\hat\varepsilon,\hat\omega)=\displaystyle{
   \frac{1}{4\pi^2}
   \big(\ln\frac{4\sqrt{2}\hat\varepsilon^{3/2}}{\hat\omega}-\gamma\big)}$\\
   &&\\
   V  & $\hat\varepsilon^{3/2}\ll\hat\omega\ll 1$ &
   $G(\hat\varepsilon,\hat\omega)=\displaystyle\frac{1}{4\pi\sqrt{3}}$
   \end{tabular}
   \end{center}
   \end{ruledtabular}
   \end{table}
%
%
   \begin{figure}[t]
   \begin{center}
   \includegraphics[width=\columnwidth]{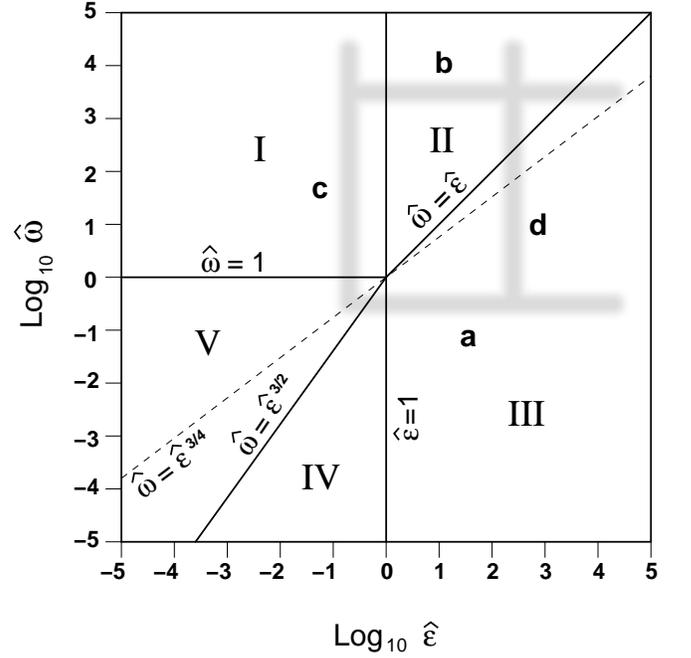}
   \end{center}
   \caption{\label{eps-omg-plane}
   The $\hat\varepsilon,\hat\omega$ plane. 
   Roman numerals from I to V mark the five regions, in which asymptotic
   limits exist. 
   Black solid lines refer to the borders between these regions.
   The gray-scale lines roughly indicate the parameter region along which 
   asymptotic formulas are compared with exact results in Fig.~\ref{Fig2} 
   (horizontal lines a and b) and in Fig.~\ref{Fig3} (vertical lines c and b).
   The broken line ($\hat \omega=\hat\varepsilon^{3/4}$) locates the lower
   boundaries $\hat \omega_p$ and $\hat \varepsilon_F$ for different densities;
   see text for more explanation.} 
   \end{figure}
%
%
\subsection{The kernel $G(\hat{\varepsilon},\hat{\omega})$.}

Results of the numerical evaluation of $G(\hat{\varepsilon},\hat{\omega})$
are depicted in Fig.~\ref{Fig1}.
One observes that the solution curves become straight equidistant lines
in the corner regions of Fig.~\ref{Fig1}.
These regions correspond to approximate asymptotic formulas of logarithmic
type in the upper tableau (linear-logarithmic scales for $G>0.1$) and of
power-law type in the lower tableau (log-log scales for $G<0.1$).
There are 5 distinct regions, indicated by Roman numbers in Fig. 1.
Their location in the $\hat{\varepsilon},\hat{\omega}$ plane 
is shown in Fig.~\ref{eps-omg-plane}.
The corresponding asymptotic expressions are given in Table 1;
they will be derived in Appendix B. 
 
The approximate expression, 
most relevant in the context of this paper, is found
for $(\hat\varepsilon+\hat\omega)\gg 1$, be it for
$\hat\varepsilon\gg 1$ or $\hat\omega\gg 1$ or both:
   \begin{equation}\label{classapprox}
   G(\hat\varepsilon,\hat\omega)
   \simeq\frac{1}{2\pi}
   \frac{1/\sqrt{2\hat\varepsilon}}
   {1-e^{-2\pi/\sqrt{2\hat\varepsilon}}}
   \ln\frac{\sqrt{\hat\varepsilon+\hat\omega}+\sqrt{\hat\varepsilon}}
   {\sqrt{\hat\varepsilon+\hat\omega}-\sqrt{\hat\varepsilon}};
   \end{equation}
it reduces to the three sublimits for cases I, II, and III in the
appropriate regions. Cases IV and V satisfy
$(\hat\varepsilon+\hat\omega)\ll 1$
and refer to the quasi-classical region: 
case IV leads to Bohr's classical Coulomb logarithm, 
and case V reproduces Kramers's classical cross-section for
bremsstrahlung (see Eq.~(\ref{Kramers})). 
Actually, the whole region V
in Fig.~\ref{eps-omg-plane} maps into essentially a horizontal line 
at $G\approx 1/(4\pi\sqrt{3})$ in Fig.~\ref{Fig1}.
The Gaunt factor defined by Eq.~(\ref{Gaunt factor})
becomes $g_{\rm ff}=1$ for case V.

For intermediate values, close to  $\hat{\varepsilon}=1$ and
$\hat{\omega}=1$ and relevant for WDM and soft X-ray photons,
the asymptotic regions  connect smoothly.
These transitional regions require full numerical evaluation.
Notice that, within the present approach, no ad hoc assumptions about
cutting off the logarithmic expressions at certain values of their
arguments are needed.
The effective collision frequencies are obtained by folding
$G(\hat{\epsilon},\hat{\omega})$ with the Fermi distributions according
to Eq.~(\ref{nu_final}). Different regions of $G(\hat{\epsilon},\hat{\omega})$ 
are probed when varying temperature, density, and photon energy. 
The regions actually probed are bounded from below
by the conditions $\omega\geq\omega_p$ and 
$\varepsilon\geq\varepsilon_F$ 
and from above by 
$\hbar\omega\leq mc^2$ and $\varepsilon\leq mc^2$.
Here $\varepsilon_F=(\hbar^2/2m)(3\pi^2n)^{2/3}$ is the Fermi energy.
Both lower bounds, $\omega_p$ and $\varepsilon_F$, depend on density only;
in Fig.~\ref{eps-omg-plane} they are located close to the broken line defined 
by $\hat\omega=\hat\epsilon^{3/4}$.
%
%
\subsection{Temperature and frequency dependence of $\nu_{\rm{eff}}$.}

\begin{figure}[!htbp]
   \begin{center}
   \includegraphics[width=\columnwidth]{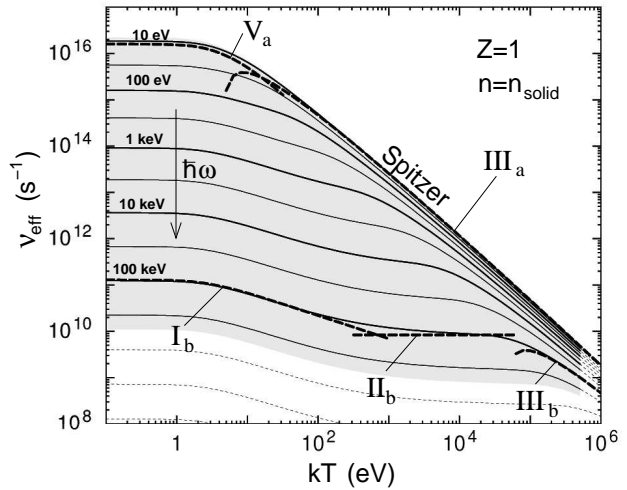}
   \end{center}
   \caption{\label{Fig2}
   The dynamic collision frequency for hydrogen ($Z=1$) at
   solid density ($n_0$), given as function of temperature $kT$ for
   fixed photon energy $\hbar\omega$, both in units of eV.
   Thick solid lines differ by factors 10.
   For sufficiently high temperature and small $\hbar\omega$, the curves
   approach the Spitzer limit, given by Eq.~(\ref{Spitzer}). 
   It is almost identical
   with the $\hbar\omega=10$~eV curve and is close to the plasma frequency 
   $\hbar\omega_p=8.3$~eV, which represents the lower frequency bound.
   Actually, the broken lines give the asymptotic formulas along the 
   horizontal lines shown in Fig.~\ref{eps-omg-plane}. 
   Line a for $\hbar\omega=10$~eV touches 
   regions III and V (marginally region IV in between), 
   and line b for $\hbar\omega=100$~keV touches regions I, II, and III. 
   One observes that the asymptotic formulas agree with the exact calculations 
   in the limited temperature regions, given in Sec. IV.   
   All curves are cut off at $kT=mc^2=511$ keV, which is the limit for the
   non-relativistic theory.
   }
   \end{figure}

\begin{figure}[!htbp]
   \begin{center}
   \includegraphics[width=\columnwidth]{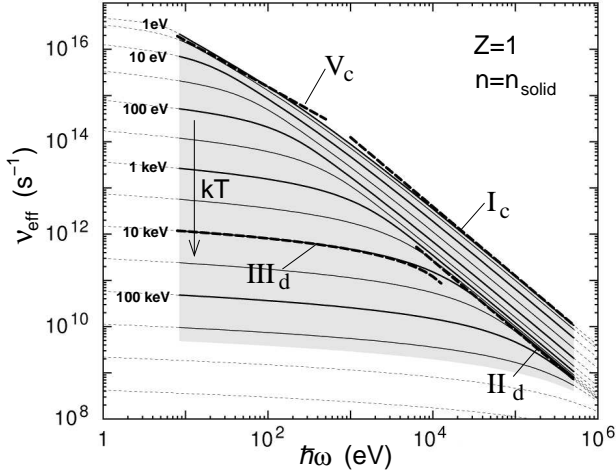}
   \end{center}
   \caption{\label{Fig3}
   The dynamic collision frequency for photon absorption in
   hydrogen ($Z=1$) at solid density ($n_0$), now plotted versus photon
   energy $\hbar\omega$ at fixed temperatures $kT$. 
   The results are shown as solid curves in the range of validity
   $\hbar\omega_p<\hbar\omega<mc^2$ and $kT<mc^2$.
   Consecutive curves differ in temperature by a factor $\sqrt{10}$.
   The broken lines refer to the approximate formulas derived in Sec. V.
   A strip of straight isotherms extends between $0\leq kT\leq \hbar\omega$.
   It scales $\propto \omega^{-3/2}$ for high photon energies 
   $\hbar\omega\gg kT$.
   The scaling becomes closer to $\nu_{\rm{eff}}\propto \omega^{-1}$ for
   small temperatures and photon energies comparable to $E_a$.
   Again Roman numerals refer to asymptotic regions specified in
   Fig.\ref{eps-omg-plane} by vertical lines; line c corresponds to $kT=1$eV
   and runs through regions I and V, while line d is for $kT=10$keV
   and extends into regions II and III.}
   \end{figure}

We have chosen hydrogen ($Z=1$) at solid density ($\rho_0=0.086$ g/cm$^3$,
$n_0=5.14\times 10^{22}$cm$^{-3}$) as a reference material.
At this density, one has $\hbar\omega_p= 8.42$~eV and 
$\varepsilon_F=kT_F=5.04$~eV.
The effective collision frequency $\nu_{\rm{eff}}$ has been calculated
numerically according to Eq.~(\ref{nu_final}) and is plotted in
Figs.~\ref{Fig2} and \ref{Fig3}.
In order to illustrate how the limiting analytic approximations 
agree with these exact numerical solutions, 
we also show them as broken lines for selected parameters 
and have labelled them with the Roman letters of the corresponding 
asymptotic regime. 
For the comparison, we have chosen the curves for 10 eV and 10 keV photons 
in Fig.~\ref{Fig2}, which probe $G(\hat\varepsilon,\hat\omega)$ 
along the horizontal lines a and b in Fig.~\ref{eps-omg-plane}, 
and in Fig.~\ref{Fig3} the 1 eV and 10 keV isotherms, probing 
$G(\hat\varepsilon,\hat\omega)$ along the vertical lines c and d, respectively.
These lines just provide 
some guiding for the path through the $\hat\varepsilon,\hat\omega$ plane, 
because $\nu_{\rm{eff}}$ actually represents an integral over $\varepsilon$, 
and we have chosen $\varepsilon\approx kT$ as an average value for this 
purpose.

In Fig.~\ref{Fig2} the collision frequency is plotted as function of 
temperature for different photon energies. 
The most prominent feature is that Spitzer's formula Eq.~(\ref{Spitzer}) 
provides the upper limit to  $\nu_{\rm{eff}}$ for sufficiently high 
temperatures ($kT>10$ eV). 
Spitzer's result nearly coincides with the curve for 10 eV photons that is 
close to the plasmon energy $\hbar\omega_p=8.4$ eV. 
In Fig.~\ref{eps-omg-plane} it corresponds to line $a$ in region III.
For low temperatures with electron energies $\hat\varepsilon<1$, 
the Spitzer formula is not valid any more. 
Here the curve for 10 eV photons saturates at a level of 
$\nu_{\rm{eff}}\approx 2\times 10^{16}$ s$^{-1}$, 
as described by Eq.~(\ref{maxabs}) obtained for region V. 
One may notice that line $a$, before stretching out to region V, 
also marginally touches region IV.
For higher photon energies $\hbar\omega>10$ eV, $\nu_{\rm{eff}}$
quickly drops.
At 100 keV photon energy ($\hat\omega\approx 4\times10^3$), 
asymptotic regions I, II, and III (compare line $b$ in 
Fig.~\ref{eps-omg-plane}) are involved, and formulas Eq.~
(\ref{nu_eff regionI}),
Eq.~(\ref{nu_eff regionII}), and Eq.~(\ref{nu_eff regionIII}) 
are seen to agree with the exact results in the corresponding regions.

In Fig.~\ref{Fig3}, the collision frequency is given in terms of isotherms 
for different photon energies. 
Here the conspicuous feature is the strip of straight-line isotherms in 
the temperature range 
$0\leq kT \leq \hbar\omega$, scaling with frequency $\propto \omega^{-3/2}$. 
This behaviour is characteristic for regions I and II and changes rapidly 
when crossing into region III at higher temperatures 
(compare Fig.~\ref{eps-omg-plane}). 
The $\omega^{-3/2}$ scaling leads to the free-free opacity scaling
$\kappa= \nu_{\rm eff}(\omega_p/\omega)^2/(\rho c)\propto \omega^{-7/2}$.
It is used in astrophysics (see e.g. Chandrasekhar \cite{Chandra}) since the
pioneering work of Gaunt in 1930 \cite{Gaunt1930}.
The present analysis reveals additional structure, unexplored
experimentally so far.
More details on the density scaling are given in the next subsection.
%
%
%
%
%
%
   \begin{figure}[!htbp]
   \begin{center}
   \includegraphics[width=\columnwidth]{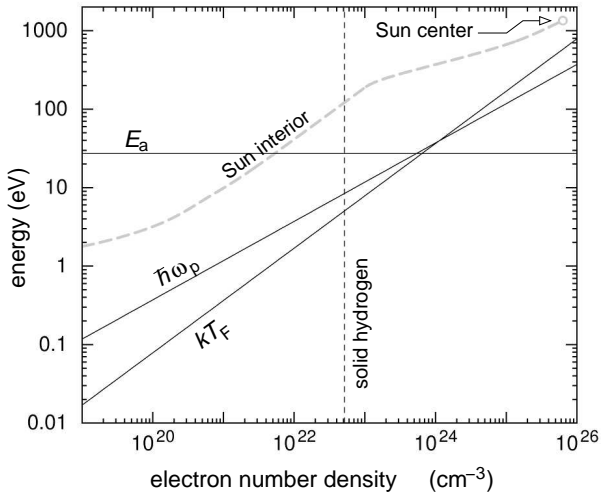}
   \end{center}
   \caption{\label{Fig4}
   Fermi energy $kT_F$,
   plasmon energy $\hbar\omega_p$, and atomic energy unit $E_A$ plotted as
   function of electron number density $n$.
   The dashed line marks the density $n_0$ of solid hydrogen.
   For comparison, also the temperature inside the sun is given versus density  
   \cite{Christensen}.}
   \end{figure}
%
%
\subsection{Density dependence of $\nu_{\rm{eff}}(\omega)$}

The effective collision frequency derived from Eq.~(\ref{nu_final})
depends on plasma density in a nonlinear way.
This is due to the Fermi temperature which scales with density like
$kT_F=\varepsilon_F\propto n^{2/3}$.
This is illustrated in Fig.~\ref{Fig4} together with the density
dependence of $\omega_p\propto n^{1/2}$ and the atomic energy unit $E_a$.
The three straight lines cut each other at a density of almost
$10^{24}$cm$^{-3}$, independently of $Z$. This is a factor 20 larger 
than the electron density $n_0$ of solid hydrogen. The three quantities
are related by 
\begin{equation}
   \frac{\hat\omega^2_pZ}{\hat T_F^{3/2}}=
   \frac{(\hbar\omega_p)^2}{(k_BT_F)^{3/2}(E_A)^{1/2}}=\frac{8\sqrt{2}}{3\pi}.
   \end{equation}

Laser plasmas generated by optical light are an example for densities 
typically below $10^{21}$cm$^{3}$, where $kT_F \ll \hbar\omega_p \ll E_a$ 
and Fermi degeneracy plays no role.
On the other hand, applications to inertial confinement fusion require highly
compressed hydrogen fuel with densities up to $10^{26}$cm$^{-3}$, 
and one has $E_a\ll\hbar\omega_p\ll kT_F $.
Here the electrons at the Fermi edge have energies of 1 keV, and 
photons of 0.5 - 1 keV energy propagate in degenerate plasma.
This implies that mainly regions II and III in the right half
of Fig.~\ref{Fig1} contribute to the integral Eq.~(\ref{nu_final}).
Densities occuring in the interior of the sun are also depicted 
in Fig.~\ref{Fig4}, for comparison; they extend over the whole density range 
and are located well above $kT_F$ in the non-degenerate regime.

\begin{figure}[!htbp]
   \begin{center}
   \includegraphics[width=\columnwidth]{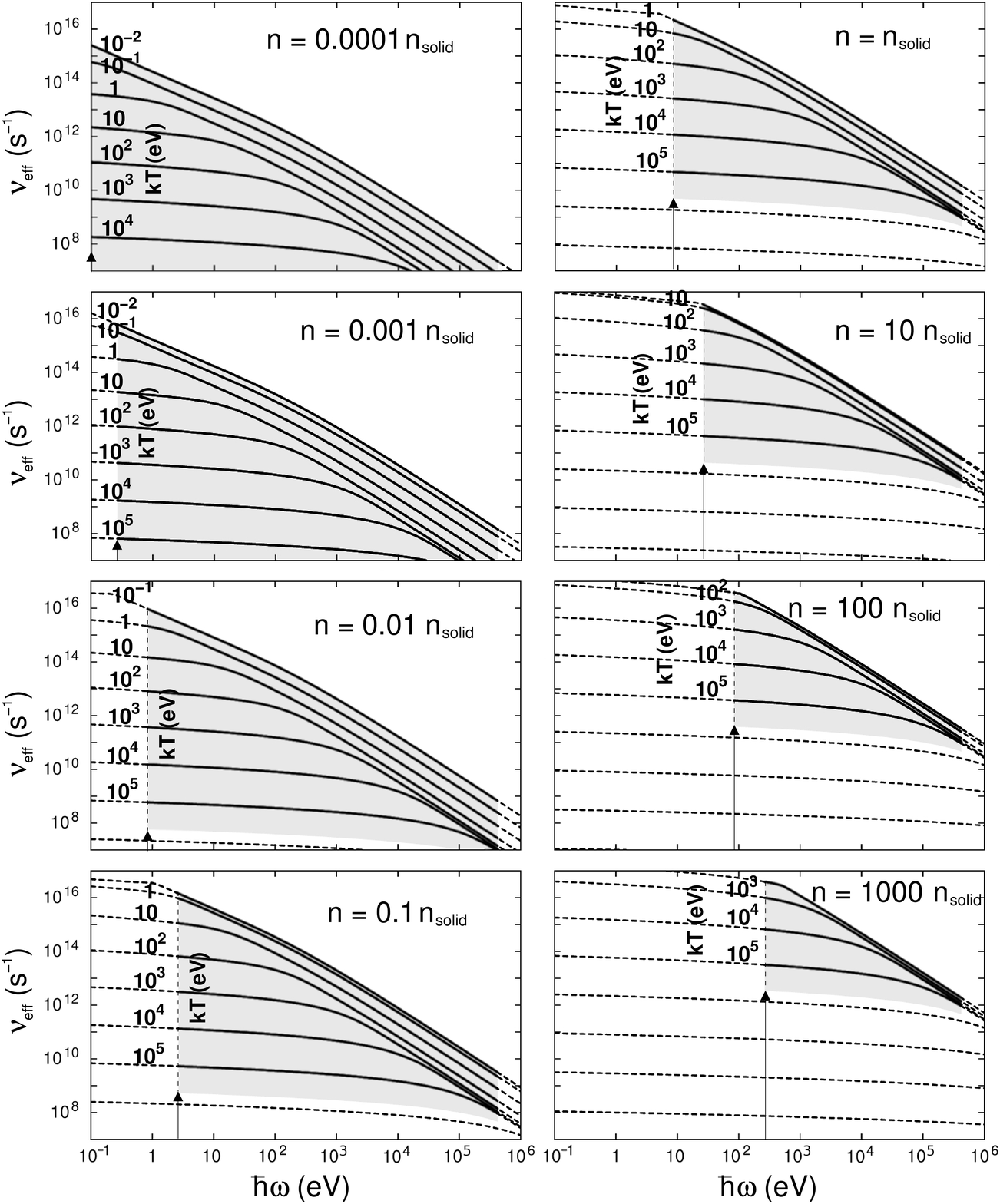}
   \end{center}
   \caption{\label{Fig5}
   Same as Fig.~\ref{Fig3}, but varying density from $10^{-4}n_0$
   to $10^{3}n_0$.
   The black arrows mark the plasmon energy $\hbar\omega_p$.
   The shaded areas show the area of validity of the present theory;
   it is limited to non-relativistic temperatures $kT\leq mc^2$
   and photon energies $\hbar\omega\leq mc^2$.}
   \end{figure}

Full numerical results for a large range of densities are shown in
Fig.~\ref{Fig5}.
The cut-off energy $\hbar\omega_p$ for photon propagation 
(shown by arrows) shifts to the right with increasing density.
The strip of isotherms showing $\omega^{-3/2}$ scaling 
decreases in width for increasing $n$.
This is because the lower boundary scales like
$\nu_{\rm{eff}}\propto (T_F/\omega)^{3/2}\propto n$, while the upper bound 
obeys $\nu_{\rm eff}\propto T_F/\omega^{3/2}\propto n^{2/3}$
(compare Eqs.(\ref{nu_eff WDM}) and (\ref{nu_eff regionII}) in Sec. IV).
One may also notice that, for the low-temperature isotherms, the transition 
from  $\omega^{-3/2}$ to quasi-classical $\omega^{-1}$ scaling is clearly
visible only for very low densities.
At very high densities, a kink is seen in the uppermost isotherms
corresponding to degenerate plasma $(kT\ll kT_F)$. 
This kink appears when $\hbar\omega$ falls below $kT_F$; 
then only a fraction of target electrons with energies 
$\varepsilon>kT_F-\hbar\omega$
can contribute to photon absorption.

Let us finally discuss the upper limit of  $\nu_{\rm{eff}}$ in Fig.~\ref{Fig5}.
Apparently, it occurs for $T\rightarrow 0$ and $\omega\rightarrow\omega_p$,
   \begin{equation}
   \nu_{\rm eff}(n,T,\omega)\le\nu_{\rm eff}(n,0,\omega_p(n))
   \equiv\nu_{\rm eff}^{\rm max}(n).
   \end{equation}
There are two limiting cases, depending on density (see Fig.~\ref{Fig4}).
For $\hat T_F\ll \hat\omega_p \ll 1$, corresponding to $n\ll 10^{24}$cm$^{-3}$
and asymptotic region V, Eq.~(\ref{maxabs}) sets the maximum:
   \begin{equation}
   \nu_{\rm{eff}}^{\rm{max}}(n)\simeq\frac{4}{3\sqrt{3}}Z\nu_0\frac{kT_F}
   {\hbar\omega_p}\propto n^{1/6}.
   \end{equation}
On the other hand, for $1\ll \hat\omega_p\ll\hat T_F$, 
corresponding to $n\gg 10^{24}$cm$^{-3}$ and asymptotic region III, 
we find from Eq.~(\ref{nu_eff III high-density})
   \begin{equation}
   \nu_{\rm{eff}}^{\rm{max}}(n)=\frac{4}{3\pi}Z\nu_0\ln\frac{4kT_F}
   {\hbar\omega_p}\propto \ln n.
   \end{equation}
%
%
%
%
%
\subsection{Z-scaling of $\nu_{\rm{eff}}$}

The dependence of  the collision frequency on ion charge $Z$ is obtained
from the hydrogen results $\nu_{\rm{eff}}(\hbar\omega,kT,n;Z_{\rm H}=1)$ by
making use of the scaling relation
   \begin{equation}
   \nu_{\rm{eff}}(\hbar\omega,kT,n;Z)=Z\nu_{\rm{eff}}
   (\hbar\omega/Z^2,kT/Z^2,n/Z^3;Z_{\rm H}=1).
   \end{equation}
This follows directly from Eq.~(\ref{nu_final}): energies scale
$\propto Z^2$, electron density $\propto Z^3$ due to
Eq.~(\ref{chem potential}), and $\nu_{\rm{eff}}$ itself $\propto Z$.
Of course, this holds only under the assumption that the plasma is fully
ionized which requires sufficiently high temperature.
With this in mind, one may apply these results also approximately 
to partially ionized plasma, replacing $Z$ by an average ion charge 
$Z_{\rm{av}}$  to decribe the free-free component of the total absorption, 
and determine bound-free and bound-bound components separately.

\begin{figure}[!htbp]
   \begin{center}
   \includegraphics[width=\columnwidth]{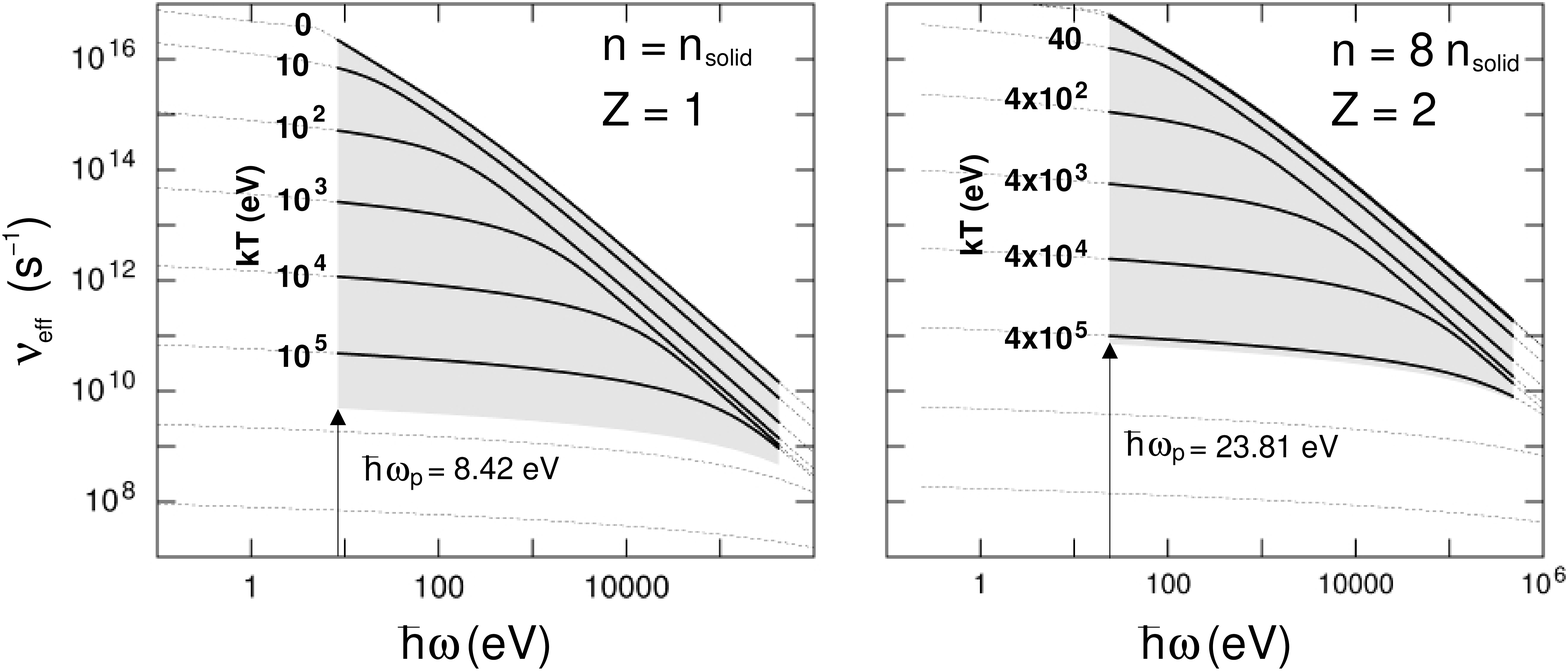}
   \end{center}
   \caption{\label{Fig6}
    Scaling of $\nu_{\rm eff}$ with ion charge $Z$. Left:
    for $Z=1$ and solid density;
   right: $\nu_{\rm{eff}}$ for $Z=2$ with scaled temperature ($kT\propto Z^2$), 
   density ($n\propto Z^3$), and plasmon energy 
   ($\hbar\omega_p\propto n^{1/2}\propto Z^{3/2}$).}
   \end{figure}

Z-scaling is illustrated in Fig.~\ref{Fig6}.
In double logarithmic presentation, the plot pattern simply shifts with $Z$.
Of course, the boundaries of validity vary differently:
the relativistic limit for $\hbar\omega<mc^2$ and $kT<mc^2$ is invariant,
and the lower limit on frequency varies according to
$\hbar\omega_p\propto n^{1/2}\propto Z^{3/2}$.
%
%
%
%
\section{Limiting regions and corresponding analytical formulas}

The general formula Eq.~(\ref{nu_final}) for $\nu_{\rm eff}$ comprises a 
number of asymptotic regions, admitting various approximate formulas.
The complexity arises because of the numerous different combinations
of the external parameters: the photon frequency $\omega$, 
the temperature $T$, and the electron density $n$.
While $n$ is a simple scaling parameter for non-degenerate
high-temperature plasma (the cases studied in most of the astrophysical
literature), the focus of the present work is on WDM at any degree of
Fermi degeneracy, and therefore $n$ enters through the Fermi temperature
$kT_F=(\hbar^2/2m)(3\pi^2n)^{2/3}$ in a non-linear way and needs to be
considered explicitly.
In addition, we have to distinguish between slow
($\varepsilon<Z^2E_a$) and fast ($\varepsilon>Z^2E_a$) electrons, and
this leads to the transitions from quasi-classical to quantum-mechanical
behaviour seen in Fig. \ref{Fig1}.
It is important for WDM. In the following, we discuss some of these cases, 
in particular to explain features seen in Figs.~\ref{Fig2} and \ref{Fig3}.
Concerning notation, we shall use both: normalized parameters
$\hat\varepsilon=\varepsilon/(Z^2E_a)$, $\hat\omega=\omega/(Z^2E_a)$,
$\hat\mu=\mu/(Z^2E_a)$, $\hat T=kT/(Z^2E_a)$,
$\hat T_F=kT_F/(Z^2E_a)=(3\pi^2na_B^3)^{2/3}/(2Z^2)$, and also the 
corresponding dimensional ones, when convenient.
%
%
\subsection{The Fermi degenerate electron gas ($kT\ll kT_F\sim\varepsilon$)}

Concerning the target electron gas, the two temperature regions
$kT\simeq 0$ and $kT\rightarrow \infty$ represent limiting situations.
At zero temperature, the Fermi function takes the shape of the step function
   \begin{equation}
   f(p)=\left\{\begin{array}{ll}1&(p<p_F)\\0&(p>p_F)\\\end{array}\right.
   \end{equation}
extending at value unity from 0 to Fermi momentum
   \begin{equation}
   p_F=\hbar(3\pi^2 n)^{1/3}
   \end{equation}
and vanishing elsewhere.
The chemical potential then becomes
   \begin{equation}\label{T_F}
   \mu=\varepsilon_F=kT_F=\frac{p_F^2}{2m}=\frac{\hbar^2}{2m}(3\pi^2n)^{2/3},
   \end{equation}
defining the Fermi energy $\varepsilon_F$ and the Fermi temperature $T_F$.

For temperatures larger than zero, the density integral 
Eq.~(\ref{chem potential}) determining $\mu$ has to be inverted.
The approximate solution
   \begin{equation} \label{Ichimaru appr}
   \frac{\mu}{kT}\simeq -\frac{3}{2}\ln\Theta+\ln\frac{4}{3\sqrt{\pi}}+
   \frac{P_1\Theta^{-(p_0+1)}+
   P_2\Theta^{-(p_0+1)/2}}{1+P_1\Theta^{-p_0}}
   \end{equation}
was given by Ichimaru \cite{Ichimaru}.
Here $\Theta=T/T_F$, and the parameters $P_1=0.25054$, $P_2=0.072$,
and $p_0=0.858$ have been adjusted to the exact result. The agreement
with the exact result is better than 0.2\%.
%
%
\subsection{The high-temperature classical plasma 
($kT_F\ll kT\sim\varepsilon$)}

For temperatures much higher than the Fermi temperature, $T\gg T_F$,
the distribution function approaches the Maxwellian distribution
   \begin{equation}\label{Maxwellian}
   f\bigg(\frac{\varepsilon-\mu}{kT}\bigg)=\frac{1}{e^{(\varepsilon-\mu)/kT}+1}
   \approx \frac{4}{3\sqrt{\pi}}\left(\frac{T_F}{T}\right)^{3/2}
   e^{-\varepsilon/kT}.
   \end{equation}
For $T\rightarrow \infty$, the third term on the right-hand side of
Eq.~(\ref{Ichimaru appr}) vanishes, while the first two give the result
Eq.~(\ref{Maxwellian}).

High temperatures imply fast target electrons ($\hat\varepsilon>1$), and we
find from Eq.~(\ref{classapprox})
   \begin{equation}\label{G fast electrons}
   G(\hat{\epsilon},\hat{\omega})\simeq \frac{1}{4\pi^2}
   \ln\frac{\sqrt{\hat\varepsilon+\hat\omega}+\sqrt{\hat\varepsilon}}
   {\sqrt{\hat\varepsilon+\hat\omega}-\sqrt{\hat\varepsilon}}.
   \end{equation}
Inserting Eqs.~(\ref{Maxwellian}) and (\ref{G fast electrons}) into
the integral (\ref{nu_final}), one obtains
   \begin{equation}\label{nu_eff high T}
   \nu_{\rm{eff}}=\frac{16}{9\pi^{3/2}}
   Z\nu_0\bigg(\frac{T_F}{T}\bigg)^{3/2} L,
   \end{equation}
where
   \begin{equation}\label{GCLog}
   L=\frac{(1-e^{-\zeta})}{\zeta}\int_0^{\infty}
   \ln\bigg(\frac{\sqrt{x+\zeta}+\sqrt{x}}{\sqrt{x+\zeta}-\sqrt{x}}\bigg)
   e^{-x}dx$$$$=\frac{e^{\zeta/2}-e^{-\zeta/2}}{\zeta}K_0(\zeta/2);
   \end{equation}
here $x=\varepsilon/kT$, $\zeta=\hbar\omega/kT$, and $K_0$ is the 
modified Bessel function of second kind and order zero.
%
%
\subsection{Asymptotic limits in region I}
Region I refers to 'slow' target electrons ($\hat\varepsilon\ll 1$) and
high-energy photons ($\hat\omega\gg 1$). 
It contributes to $\nu_{\rm{eff}}$ in Eq.~(\ref{nu_final}) only for 
sufficiently
small density such that $\hat T_F\ll 1$; this is the case e.g. for hydrogen
plasma at solid density.
In region I, one has $G(\hat\varepsilon,\hat\omega)\approx
(\pi\sqrt{2\hat\omega})^{-1}$ (compare Table~\ref{table}), and 
the integral Eq.~(\ref{nu_final}) can be solved analytically. 
Making use of
   \begin{equation}
   \int_0^{\infty}\frac{dx}{1+e^{x+a}}=\ln(1+e^{-a}),
   \end{equation}
we find
   \begin{equation}\label{nu_eff WDM}
   \nu_{\rm{eff}}\simeq\frac{16\pi}{3}Z\nu_0
   \frac{1}{\pi\sqrt{2\hat\omega}}\frac{\hat T}{\hat\omega}
   \ln\frac{1+e^{\hat\mu/\hat T}}{1+e^{(\hat\mu-\hat\omega)/\hat T}}.
   \end{equation}
Here the denominator $(1+\exp{((\hat \mu-\hat\omega)/\hat T)}$ under the
logarithm in Eq.~(\ref{nu_eff WDM}) approaches unity for $\hat\omega\gg 1$
so that we have
   \begin{equation}\label{nu_eff regionI} 
   \nu_{\rm eff} \simeq \frac{16}{3\sqrt{2}} Z\nu_0
   \frac{\hat T}{\hat\omega^{3/2}}
   \ln{(1+\exp(\hat\mu/\hat T))};
   \end{equation}
Combined with Eq.~(\ref{Ichimaru appr}) for $\mu$,
this covers region I for any value of the degeneracy parameter $\Theta=T/T_F$. 
A comparison with the full numerical solution is given in Fig.~\ref{Fig2} 
for 100 keV photons.
For $T\rightarrow 0$, Eq.~(\ref{nu_eff regionI}) becomes
   \begin{equation}\label{nu_eff WDM} 
   \nu_{\rm{eff}} = \frac{16}{3\sqrt{2}} Z\nu_0 
   \frac{\hat T_F}{\hat\omega^{3/2}}
   \end{equation}
and describes the upper boundary for $\nu_{\rm{eff}}$ for sufficiently 
high photon energies in Fig.~\ref{Fig3}.
%
%
\subsection{Asymptotic limits in region II}
Region II refers to electron energies satisfying 
$1\ll\hat\varepsilon\ll\hat\omega$, where
$G(\hat\varepsilon,\hat\omega)
\simeq (1/2\pi^2)\sqrt{\hat\varepsilon/\hat\omega}$ applies. 
The integral in Eq.~(\ref{nu_final}) then reduces essentially to the density
integral 
$\int d\hat\varepsilon\sqrt{\hat\varepsilon}f((\hat\varepsilon-\hat\mu)/\hat T)
=(2/3)\hat T_F^{3/2}\propto n$
(see Eq.~(\ref{chem potential})), and we find
   \begin{equation}\label{nu_eff regionII} 
   \nu_{\rm{eff}} \simeq \frac{16}{9\pi} Z\nu_0
   \bigg(\frac{\hat T_F}{\hat\omega}\bigg)^{3/2}.   
   \end{equation}
Notice that $\nu_{\rm eff}$ becomes independent of temperature. 
Such a tendency is clearly observed in Fig.~\ref{Fig2} and is responsible 
for the clustering at the lower boundary of the strip of isotherms 
in Fig. \ref{Fig3}.
One may also notice 
that Eq.~(\ref{nu_eff regionII}) holds for both classical plasma 
in the range $1\ll\hat T\ll\hat \omega$ and degenerate plasma, 
occuring at high densities for $1\ll\hat T_F\ll\hat \omega$.
%
%
\subsection{Asymptotic limits in region III}

We now turn to temperatures much larger than the photon energy,
such that $\hat\omega\ll\hat\epsilon$.
In this limit, the integral Eq.~(\ref{GCLog}) becomes $L=\ln(4/\zeta)-\gamma$,
where $\gamma=0.577$ is Euler's constant, and one obtains 
   \begin{equation}\label{nu_eff regionIII}
   \nu_{\rm{eff}}=\frac{4\sqrt{2\pi}}{3}\frac{Zne^4}{\sqrt{m}(kT)^{3/2}}
   \bigg(\ln\frac{4kT}{\hbar\omega}-\gamma\bigg).
   \end{equation}
Here we have expressed the Fermi temperature Eq.~(\ref{T_F}) 
in Eq.~(\ref{nu_eff high T}) by the elementary quantities
to recover Spitzer's formula Eq.~(\ref{Spitzer}) in its well known form. 
The front factor is identical to what is found in standard text books 
(see e.g. Ref. \onlinecite{Wesson}), and the logarithmic term reproduces 
the quantum Coulomb logarithm of Bethe in the limit 
$\omega\rightarrow \omega_p$. 
\footnote{One may conjecture that this limit corresponds
to the limit $\omega\rightarrow 0$, had we treated Debye screening correctly,
and one is tempted to substitute $\sqrt{\omega^2+\omega_p^2}$ for $\omega$
under the logarithm in Eq.~{\ref{nu_eff regionIII}}.}
In Fig.~\ref{Fig2}, Eq.~(\ref{nu_eff regionIII}) is seen 
to provide the upper boundary of $\nu_{\rm{eff}}$ for sufficiently 
high temperature. 

At sufficiently high density, the Fermi temperature $kT_F$ becomes larger
than the plasmon energy $\hbar\omega_p$, and light may propagate
in degenerate plasma ($1\ll\hat\omega_p\le\hat\omega\ll\hat T_F$
and $\hat T\ll\hat T_F$).
The integral Eq.~(\ref{nu_final}) is then performed
over step-like distribution functions:
   \begin{equation}
   \nu_{\rm{eff}}\simeq\frac{16\pi}{3}Z\nu_0\frac{1}{4\pi^2}
   \left[
   \int_{kT_F-\hbar\omega}^{kT_F}
   \ln\frac{\sqrt{\varepsilon+\hbar\omega}+\sqrt{\varepsilon}}
   {\sqrt{\varepsilon+\hbar\omega}-\sqrt{\varepsilon}}
   \frac{d\varepsilon}{\hbar\omega}\right];
   \end{equation}
here the lower boundary of the integral indicates that only electrons
with energy $\varepsilon>kT_F-\hbar\omega$ 
contribute to absorption due to the Pauli principle, and we find
   \begin{equation}\label{nu_eff III high-density}
   \nu_{\rm{eff}}=\frac{4}{3\pi}Z\nu_0\ln\frac{4\hat T_F}{\hat\omega}.
   \end{equation}
%
%
\subsection{Asymptotic limit for region IV}

For sufficiently low density (such that $\hat T_F\ll 1$), an additional 
asymptotic regime is found in region IV, where 
$\hat T_F\ll\hat\omega_p\le\hat\omega\ll\hat\varepsilon^{3/2}<1$ holds.
This corresponds to a non-degenerate plasma with Maxwellian electron 
distribution.
Using the approximate expression for $G(\hat\epsilon,\hat\omega)$, 
valid for region IV (see Table~\ref{table}), Eq.~(\ref{nu_final}) 
for $\nu_{\rm{eff}}$ takes the form
   \begin{equation}
   \nu_{\rm{eff}}=\frac{16}{9\pi^{3/2}}Z\nu_0
   \bigg(\frac{\hat T_F}{\hat T}\bigg)
   ^{3/2}\frac{1-e^{-{\hat\omega/\hat T}}}{\hat\omega/\hat T}
   $$$$\times
   \int_0^{\infty}
   \big(\ln\frac{4\sqrt{2}\hat\varepsilon^{3/2}}{\hat\omega}-\gamma\big)
   \exp{(-\frac{\hat\varepsilon}{\hat T})}
   \frac{d\hat\varepsilon}{\hat T}.
   \end{equation}
Recalling $\int_0^{\infty}\ln xe^{-x}dx=-\gamma$, one gets
   \begin{equation}\label{nu_eff classical Bohr}
   \nu_{\rm{eff}}=\frac{4\sqrt{2\pi}}{3}\frac{Zne^4}{\sqrt{m}(kT)^{3/2}}
   \bigg(\ln \frac{4\sqrt{2}(kT)^{3/2}}{Ze^2\omega \sqrt{m}} 
   -\frac{5}{2}\gamma\bigg). 
   \end{equation}
This result is similar to Eq.~(\ref{nu_eff regionIII}), but under
the logarithm it shows the parametric combination 
$T^{3/2}/(Z\omega)$
characteristic for Bohr's classical Coulomb logarithm in contrast to
Bethe's quantum-mechanical result $\propto \ln(4kT/\hbar\omega)$ 
(see e.g. Ref. \onlinecite{Atzeni&MtV}). 
So even the transition between the classical and quantum Coulomb logarithms
is contained in the present formalism and
naturally emerges under the appropriate limit.
For solid-density hydrogen, region VI is just marginally touched 
(compare line $a$ in Fig.~\ref{eps-omg-plane}) and does not clearly show up 
in Fig.~\ref{Fig2}.
%
%
\subsection{Asymptotic limit in region V}
%
%
This region is defined by the condition 
$\hat\varepsilon^{3/2}\ll\hat\omega\ll1$.
Here $G(\hat\varepsilon,\hat\omega)\approx 1/(4\pi\sqrt{3})$ 
(compare Table~\ref{table}) is independent of $\hat\varepsilon$ as in region I, 
and one obtains 
   \begin{equation}\label{lowT,low-omega}
   \nu_{\rm{eff}}\simeq\frac{4}{3\sqrt{3}}Z\nu_0\frac{\hat T}
   {\hat\omega}\ln\frac{1+\exp{(\hat\mu/\hat T)}}
   {1+\exp{((\hat\mu-\hat\omega)/\hat T)}}.
   \end{equation}
There exist a number of sublimits here; however, it may be best to use the
full Eq.~(\ref{lowT,low-omega}) in combination with Eq.~(\ref{Ichimaru appr}).
In the limit $\hat T\rightarrow 0$ and $\hat\omega\rightarrow \hat\omega_p\gg
T_F$,
   \begin{equation}
   \nu_{\rm{eff}}\simeq\frac{4}{3\sqrt{3}}Z\nu_0\frac{\hat T_F}{\hat \omega_p};
   \label{maxabs}
   \end{equation}
it depends only on density and scales $\propto n^{1/6}$.
It describes the maximum value of $\nu_{\rm{eff}}$ in Fig. \ref{Fig5} for
a given density well below the solid density.
%
%
\section{Summary and Outlook}
Collisional free-free photon absorption in dense plasma has been studied
with particular focus on warm dense matter. 
A global expression for the effective collision frequency 
$\nu_{\rm{eff}}(T,\hbar\omega,n)$
has been derived from Sommerfeld's cross section for bremsstrahlung. 
It holds for frequencies above the plasma frequency and in the
non-relativistic regime.
Numerical results for fully ionized hydrogen plasma at solid density
are presented 
over a wide range of temperatures $T$ and photon energies $\hbar\omega$.
Also the dependence on density $n$ is obtained over a wide range, 
covering any degree of degeneracy. For ions with charge number $Z>1$,
the relation for $Z$-scaling is derived.

We have identified a number of asymptotic regions in which the numerical 
results can be represented by simple analytical formulas. 
For sufficiently high temperatures and $\hbar\omega\ll kT$, 
Spitzer's formula for a classical plasma is reproduced 
and appears as an upper bound to $\nu_{\rm{eff}}$.
Concerning scaling with frequency, $\nu_{\rm{eff}}\propto \omega^{-3/2}$ is 
found for temperatures $kT\ll \hbar\omega$, independent of the degree of 
degeneracy; it leads to the free-free opacity scaling 
$\kappa\propto \omega^{-7/2}$, well-known in astrophysics.
For intermediate regions, relevant to warm dense matter, new asymptotic 
relations are found that should be tested experimentally.

The present results are based on the quantum-mechanical solution 
of radiant electron collisions with ions.
The most difficult part of the work has been the numerical evaluation 
of the hypergeometric functions involved. 
This problem has been investigated independently in the astrophysical 
literature, calculating the free-free Gaunt factor. 
The methods used in the present work are outlined in the Appendices.

For applications,  it is planned, in a companion paper, to provide a global 
fit formula, based on the present work and sufficiently accurate for direct 
use in numerical codes.
%
%
\section*{Acknowledgements}
R. R. acknowledges the hospitality of the Max-Planck-Institut für Quantenoptik,
where most of the present work was done, and financial support from 
the Spanish Ministerio de
Economía y Competitividad, Project No. ENE2014-54960-R, and by the
EUROfusion Consortium under the project AWP15-ENR-01/CEA-02.
J. MtV thanks Stefanie Hansen and Peter Mulser for advice concerning 
evaluations of the Gaunt factor in the astrophysical literature and also 
for discussions with Thomas Blenski on photo-absorption in dense plasma.
%
%
\appendix
%
%
\section[A]{Numerical evaluation of $G(\hat\epsilon,\hat\omega)$}
The numerical evaluation of the kernel $G(\hat\epsilon,\hat\omega)$ is 
not trivial due to the presence of hypergeometric functions.
As such functions are not available
in standard libraries for all possible combinations of large/small values
of the arguments, numerical techniques taylored for each specific 
problem are frequently required \cite{pearson}.
Here, in order to be close to literature standards
\cite{Landau&Lifshitz, KrainovII, BH1962},
we introduce a notation slightly different from the one used
in the main text, namely, $\nu\equiv \eta_+$ and 
$\nu'\equiv \eta$. 
Eq.~(\ref{G}) now takes the form
   \begin{equation}\label{eq034}
   G(\hat{\varepsilon},\hat{\omega})=\frac{\xi}{4}\frac
   {d|F(i\nu',i\nu;1;\xi)|^2/d\xi}{(1-e^{-2\pi\nu'})(e^{2\pi\nu}-1)},
   \end{equation}
where
   \begin{equation}\label{eq037}
   \nu=\frac{1}{\sqrt{2(\hat \varepsilon+\hat\omega)}},\hspace{5mm}
   \nu'=\frac{1}{\sqrt{2\hat \varepsilon}},\hspace{5mm}
   \xi=-\frac{4\nu\nu'}{(\nu-\nu')^2}.
   \end{equation}
The derivative in Eq.~(\ref{eq034})
can be expressed as (Eq.~(15.5.1) of Ref. \onlinecite{nist})
   \begin{equation}\label{eq004}
   \frac{d}{d\xi}|F(i\nu',i\nu;1;\xi)|^2=2\mathfrak{R}\left[
   F^*(i\nu',i\nu;1;\xi)\frac{d}{d\xi}F(i\nu',i\nu;1;\xi)\right],
   \end{equation}
   \begin{equation}\label{eq009}
   \frac{d}{d\xi}F(i\nu',i\nu;1;\xi)=-\nu\nu'F(1+i\nu',1+i\nu;2;\xi).
   \end{equation}
Hypergeometric functions are defined by
Gauss series (Eq.~(15.2.1) of Ref. \onlinecite{nist})
   \begin{equation}
   \label{eq001}
   F(a,b;c;\xi)=1+\frac{ab}{c}\frac{\xi}{1!}+
   \frac{a(a+1)b(b+1)}{c(c+1)}\frac{\xi^2}{2!}+\dots
   \end{equation}
on the disk $|\xi|<1$, and by analytic continuation elsewhere.
In our case, $\xi$ is always real and negative.
For $\xi\le-1$, one can use the expression (Eq.~(15.3.8) of Ref.
\onlinecite{abramovitz})
   \begin{equation}
   \label{eq002}
   F(a,b;c;\xi)=
   $$$$
   \frac{\Gamma(c)\Gamma(b-a)}{\Gamma(b)\Gamma(c-a)}(1-\xi)^{-a}
   F\left(a,c-b;a-b+1;\frac{1}{1-\xi}\right)
   $$$$
   +\frac{\Gamma(c)\Gamma(a-b)}{\Gamma(a)\Gamma(c-b)}(1-\xi)^{-b}
   F\left(b,c-a;b-a+1;\frac{1}{1-\xi}\right),
   \end{equation}
that maps the interval $\xi\in(-\infty,-1)$ into the interval $(0,1/2)$,
where the series converge. 
Gamma functions of complex arguments have been evaluated by the algorithm
described in Sec. 6.1 of Ref. \onlinecite{recipes}. 
For $-1\le\xi<0$, we use 
(Eq.~(15.3.4) of Ref. \onlinecite{abramovitz})
   \begin{equation}
   \label{eq003}
   F(a,b;c;\xi)=(1-\xi)^{-a}F\left(a,c-b;c;\frac{\xi}{\xi-1}\right),
   \end{equation}
that maps the interval $\xi\in(-1,0)$ into the interval $(0,1/2)$,
ensuring fast convergence of the series even if $\xi$ is close to $-1$.
However, severe cancellation errors prevent the use of Eqs. (\ref{eq002}) 
and (\ref{eq003}) for large values 
of the parameters (e.g., for $\nu/\nu'=1/2$ and double precision arithmetic,
Eq.~(\ref{eq002}) fails for $\nu'\gtrsim 250$).
To overpass this difficulty, arbitrary precision floating
point arithmethic was used by van Hoof et al. \cite{Hoof2014}.
Here, we apply a different approach, using the
following integral representation (Eq.~(15.6.1) of Ref. \onlinecite{nist})
   \begin{equation}
   F(a,b;c;\xi)=\frac{\Gamma(c)}{\Gamma(b)\Gamma(c-b)}
   \int_0^1\frac{t^{b-1}(1-t)^{c-b-1}}{(1-\xi t)^a}dt,$$$$
   \mathfrak{R}c>\mathfrak{R}b>0,
   \end{equation}
where, $a=i\nu'$, $b=i\nu$, and $c=1$. 
Defining $\rho=\nu/\nu'$ and using the ``reflection formula'' 
$\Gamma(z)\Gamma(1-z)={\pi}/{\sin \pi z}$ (Eq.~(5.5.3) of Ref. 
\onlinecite{nist}), 
one gets
   \begin{equation}\label{eq023}
   F(i\nu',i\rho\nu';1,\xi)=\frac{i\sinh(\pi\rho\nu')}{\pi} I,
   \end{equation}
where
   \begin{equation}
   I=\int_0^1t^{i\rho\nu'-1}(1-t)^{-i\rho\nu'}(1-\xi t)^{-i\nu'}dt
   \end{equation}
The requirement $\mathfrak{R}b>0$ is not satified,
but, because hypergeometric functions are continuous in the parameters, we
can evaluate the integral for $b=i\rho\nu'+\varsigma$ 
($\varsigma$ real and positive), and afterwards take the
limit $\varsigma\rightarrow0$.
The derivative of $F$ is obtained by taking the derivative of the integrand
of Eq.~(\ref{eq023}) with respect to $\xi$
   \begin{equation} \label{eq035}
   \frac{d}{d\xi}F(i\nu',i\rho\nu';1;\xi)
   =-\frac{\nu'\sinh(\pi\rho\nu')}{\pi}I_A,
   \end{equation}
where
   \begin{equation} \label{eq038}
   I_A=\int_0^1
   t^{i\rho\nu'}(1-t)^{-i\rho\nu'}(1-\xi t)^{-i\nu'-1}dt.
   \end{equation}
Direct application of these expressions in Eq.~(\ref{eq034}) can cause
also cancelation problems for large $\nu$ and $\nu'$.
These can be overcome by evaluating 
   \begin{equation}\label{eq014}
   I_B=I-\left(\frac{2}{1-\rho}-\xi\right)I_A=$$$$
   \int_0^1\left(\frac{1}{t}-\frac{2}{1-\rho}\right)t^{i\rho\nu'}
   (1-t)^{-i\rho\nu'}(1-\xi t)^{-i\nu'-1}dt
   \end{equation}
instead of $I$.
Finally Eq.~(\ref{eq034}) takes the form
   \begin{equation}\label{eq015}
   G=\frac{\nu'|\xi|}{16\pi^2}
   \frac{1-e^{-2\pi\nu}}{1-e^{-2\pi\nu'}}|I_AI_B^*-I_A^*I_B|.
   \end{equation}
The methods used to evaluate numerically the integrals $I_A$ and $I_B$ are 
described in Appendix C.
%
%
\section[B]{Asymptotic limits of $G(\hat\varepsilon,\hat\omega)$}

The asymptotic limits of the Sommerfeld cross section have been widely
studied \cite{Sommerfeld,Landau&Lifshitz,KrainovI,KrainovII,shkarofsky}. 
However, the derivations of particular cases are rather scattered through 
the literature, frequently using different notations and normalization factors.
For this reason, we introduce in this appendix 
a systematic derivation of the five asymptotic limits displayed
in table 1.
It should be mentioned that we could not find, in the literature,
an asymptotic limit valid for the full domain III-IV,
similar to the one derived below.
As indicated in Appendix A, the hypergeometric 
functions that appear in $G(\hat\varepsilon,\hat\omega)$ can be evaluated
in three different ways, namely either using Eq.~(\ref{eq001}), 
Eq.~(\ref{eq002}), or Eq.~(\ref{eq023}).
In one to one correspondence with these three ways,
three asymptotic limits for
$G(\hat\varepsilon,\hat\omega)$ have been derived.
%
%
\subsection{Regions I and II}
If $|\xi|<1$ and $\nu\ll1$,  
the power expansion in Eq.~(\ref{eq009}) simplifies to
   \begin{equation}\label{eq011}
   \frac{d}{d\xi}F(i\nu',i\nu;1;\xi)=$$$$
   -\nu\nu'F(1+i\nu',1+i\nu;2;\xi)\simeq
   -\nu\nu'F(1+i\nu',1;2;\xi)=$$$$
   -\nu\nu'\left(1+\frac{(1+i\nu')}{2}\frac{\xi}{1!}+
   \frac{(1+i\nu')(2+i\nu')}{3}\frac{\xi^2}{2!}+\dots\right).
   \end{equation}
Notice that if $\nu'\sim\nu\ll1$, the simplified series becomes
$\sum_{n=0}^{\infty}\xi^n/(1+n)=-\ln(1-\xi)/\xi$.
On the other hand, if $\nu'\gg\nu$, then
$|\xi|=4\nu\nu'/(\nu-\nu')^2\simeq4\nu/\nu'\ll 1$, and
all $\xi$ power terms in Eq.~(\ref{eq011}) can be neglected, so that 
$F(1+i\nu',1;2;\xi)\simeq 1=-\lim_{\xi\rightarrow 0}\ln(1-\xi)/\xi$.
This leads in both cases to
   \begin{equation}
   \frac{d}{d\xi}F(i\nu',i\nu;1;\xi) \simeq
   \nu\nu'\frac{\ln(1-\xi)}{\xi}.
   \end{equation}
The hypergeometric function itself can be evaluated by integration 
   \begin{equation}\label{eq012}
   F(i\nu',i\nu;1;\xi)=F(i\nu',i\nu;1;0)+\int_0^\xi
   \frac{d}{d\xi}F(i\nu',i\nu;1;\xi)d\xi\simeq$$$$
    1+\nu\nu'\int_0^\xi\frac{\ln(1-\xi)}{\xi}d\xi\simeq 1.
   \end{equation}
For $|\xi|<1$, the integrand is bounded $|\ln(1-\xi)/\xi|\le1$, 
so that the absolute value of the second term in Eq.~(\ref{eq012}) is less that 
$|\nu\nu'\xi|\sim \nu^2\ll1$ and can be neglected.
Using Eqs.~(\ref{eq011}) and (\ref{eq012}) in Eq.~(\ref{eq034}), one gets
   \begin{equation}\label{eq026}
   G\simeq\frac{1}{2\pi}\frac{\nu'}{1-e^{-2\pi\nu'}}
   \ln\frac{\nu'+\nu}{\nu'-\nu}=$$$$
   \frac{1}{2\pi}\frac{1/\sqrt{2\hat\varepsilon}}
   {1-e^{-2\pi/\sqrt{2\hat\varepsilon}}}
   \ln\frac{\sqrt{\hat\varepsilon+\hat\omega}+\sqrt{\hat\varepsilon}}
   {\sqrt{\hat\varepsilon+\hat\omega}-\sqrt{\hat\varepsilon}}.
   \end{equation}
In terms of $\hat\varepsilon$ and $\hat\omega$, 
the conditions  $\nu\ll1$ and $|\xi|<1$
become $\hat\varepsilon+\hat\omega\gg1$ and
$\hat\omega>4(4+3\sqrt2)\hat\varepsilon$, respectively. 
So  Eq.~(\ref{eq026}) is valid deeply inside the
joint regions marked as I and II in Fig. \ref{eps-omg-plane}.
%
%
%
\subsection{Regions III and IV}
For $|\xi|>1$, the Taylor series in Eqs. (\ref{eq001}) 
diverges, and one must use Eq.~(\ref{eq002}).
In that case, it proves convenient to introduce the auxiliar variables
$\epsilon=\frac{1}{2}(\nu'-\nu)$, $\bar\nu=\frac{1}{2}(\nu'+\nu)$.
Here, we assume that $\epsilon\ll1$ and $\epsilon\ll\bar \nu$
(although $\bar\nu$ can be arbitrary).
With $a=i(\bar\nu+\epsilon)$, $b=i(\bar\nu-\epsilon)$,
and $c=1$, Eq.~(\ref{eq002}) takes the form 
   \begin{equation}\label{eq029}
   F(i\nu',i\nu;1;\xi)=
   \frac{\Gamma(-2i\epsilon)}{\Gamma(i(\bar \nu-\epsilon))
   \Gamma(1-i(\bar\nu+\epsilon))}(1-\xi)^{-i(\bar \nu+\epsilon)}$$$$\times
   F\left(i(\bar \nu+\epsilon),1-i(\bar \nu-\epsilon);1+2i\epsilon;
   \frac{1}{1-\xi}\right)$$
   $$+\frac{\Gamma(2i\epsilon)}{\Gamma(i(\bar \nu+\epsilon))
   \Gamma(1-i(\bar\nu-\epsilon))}(1-\xi)^{-i(\bar \nu-\epsilon)}$$$$\times
   F\left(i(\bar \nu-\epsilon),1-i(\bar \nu+\epsilon);1-2i\epsilon;
   \frac{1}{1-\xi}\right).
   \end{equation}
Due to $\Gamma(0)=\infty$, the two terms are singular for 
$\epsilon\rightarrow 0$ with opposite sign.
Thus, to evaluate this expression for $\epsilon\ll 1$, it is neccessary 
to expand each term in powers of $\epsilon$ up to first order.
The hypergeometric functions, having a small argument
$1/(1-\xi)=\epsilon^2/\bar \nu^2$, can be expressed as fast convergent series
   \begin{equation}
   F(i(\bar \nu\pm\epsilon),1-i(\bar \nu\mp\epsilon),1\pm2i\epsilon,
   \frac{\epsilon^2}{\bar \nu^2})=$$$$
   1+\frac{i(\bar\nu\pm\epsilon)(1-i(\bar\nu\mp\epsilon))}
   {1\pm2i\epsilon}\frac{\epsilon^2}{\bar\nu^2}+\dots
   \end{equation}
As $\epsilon\ll\min(1,\bar\nu)$, the leading term of the series 
is either of order $\epsilon^2$, if $\bar\nu\gg1$, or of order
$\epsilon^2/\bar\nu$, if $\bar\nu\ll1$. 
In both cases, it can be neglected in comparison
with terms of order $\epsilon$, and one may set $F=1$ in 
the right hand side of Eq.~(\ref{eq029}).
The factors containing the gamma functions can be simplified 
using the identities (Eqs. (5.4.1), (5.5.1), and (5.5.3) of Ref. 
\onlinecite{nist}) 
$\Gamma(1)=1$, 
$z\Gamma(z)=\Gamma(z+1)$,
$\Gamma(z)\Gamma(1-z)=\pi/\sin(\pi z)$, $\sin(iz)=i\sinh(z)$,
and the Taylor expansion 
   \begin{equation}
   \Gamma(z+\Delta z)=\Gamma(z)+\Gamma'(z)\Delta z+\dots=
   \Gamma(z)(1+\psi(z)\Delta z+\dots),
   \end{equation}
where $\psi(z)\equiv\Gamma'(z)/\Gamma(z)$ is the psi (digamma) 
function (Eq.~(5.2.2) of Ref.  \onlinecite{nist}).
Retaining terms up to first order in $\epsilon$;
   \begin{equation}
   F(i\nu',i\nu;1;\xi)\simeq
   \frac{\bar\nu}{\epsilon}\frac{\sinh(\pi\bar\nu)}{2\pi}
   (1-\xi)^{-i\bar \nu}$$$$\times
   \Big(-\frac{1+ 2i\epsilon a(\bar\nu)}
   {\bar\nu+\epsilon}
   (1-\xi)^{-i\epsilon}
   +\frac{1- 2i\epsilon a(\bar\nu)}
   {\bar\nu-\epsilon}
   (1-\xi)^{i\epsilon}\Big),
   \end{equation}
where we introduce the function
$a(\bar\nu)\equiv-\psi(1)+\frac{1}{2}(\psi(i\bar\nu)+\psi(-i\bar\nu))$.
From this expression, it is straightforward to obtain
   \begin{equation}
   \frac{d|F|^2}{d\xi}\simeq
   \frac{\sinh^2(\pi\bar\nu)}{\pi^2(\bar\nu^2-\epsilon^2)}
   \epsilon(4\epsilon a(\bar\nu)\cos(2\epsilon\ln(1-\xi))$$$$-
    (1-4\epsilon^2a(\bar\nu)^2)\sin(2\epsilon\ln(1-\xi))),
   \end{equation}
and, taking into account that $\epsilon\ll\bar\nu\simeq\nu\simeq\nu'$
and $\epsilon\ll 1$,
   \begin{equation}
   \frac{d|F|^2}{d\xi}\simeq\frac{4\sinh^2(\pi\nu')}{\pi^2\xi}
   (\ln\frac{\nu'}{\epsilon}-a(\nu')).
   \end{equation}
Finally, using this expression in Eq.~(\ref{eq034}), one gets
   \begin{equation}\label{eq008}
   G\simeq\frac{1}{4\pi^2}(\ln\frac{2\nu'}{\nu'-\nu}
   -a(\nu'))
   \simeq \frac{1}{4\pi^2}(\ln\frac{4\hat\varepsilon}{\hat\omega}
   -a(1/\sqrt{2\hat\varepsilon})).
   \end{equation}
Although function $a(\bar \nu)$ contains the special function $\psi(\bar\nu)$,
using Eq.~(5.7.6) of Ref. \onlinecite{nist}, it can be transformed 
into a form suitable for efficient numerical evaluation
   \begin{equation}
   a(\bar\nu)=\bar\nu^2\sum_{n=1}^{\infty}\frac{1}{n(n^2+\bar\nu^2)}
   \simeq 
   $$$$\bar\nu^2\sum_{n=1}^{N}\frac{1}{n(n^2+\bar\nu^2)}+
   \frac{1}{2}\ln\left(\frac{\bar\nu^2}{(N+1/2)^2}+1\right).
   \end{equation}
For $N=100$,  $a(\bar\nu)$ is obtained with a relative error less than 
$10^{-6}$ for any positive value of $\bar\nu$.
For small and large arguments, 
using Eqs. (5.4.12), (5.11.2), and (25.2.1) of Ref. \onlinecite{nist}, one has
   \begin{equation}
   a(\bar \nu)\simeq\left\{\begin{array}{ll}
   \zeta(3)\bar\nu^2 & \text{for $\bar\nu\ll1$}\\
   &\\
   \gamma+\ln \bar\nu & \text{for $\bar\nu\gg1$}\\
   \end{array}\right.
   \end{equation}
where $\zeta(3)=1.20205\dots$ is Apéry's constant ($\zeta$ is the Riemann Zeta 
function) and $\gamma=0.57721\dots$ is Euler's constant.
Eq.~(\ref{eq008}) has been derived under the assumption 
$\epsilon\ll\min(1,\bar \nu)$, which can be written as
   \begin{equation}\label{eq028}
   \sqrt{\hat\varepsilon+\hat\omega}-\sqrt{\hat\varepsilon}\ll\min(
   \sqrt{\hat\varepsilon}\sqrt{\hat\varepsilon+\hat\omega},
   \sqrt{\hat\varepsilon+\hat\omega}+\sqrt{\hat\varepsilon}).
   \end{equation}
This is only possible if $\hat\omega\ll\hat\varepsilon$.
In that case, Eq.~(\ref{eq028}) becomes $\hat\omega\ll\min(\hat\varepsilon,
\hat\varepsilon^{3/2})$, and it is satisfied in the
regions marked as III and IV in Fig. \ref{eps-omg-plane}.
%
%

\subsection{Region V}
This situation corresponds to the quasi-classical limit and has been
treated in \S92 of Ref. \onlinecite{Landau&Lifshitz}.
Here we obtain the asymptotic limit from Appendix A.
The integrals defined by Eqs. (\ref{eq038}) and (\ref{eq014}) can be written as 
   \begin{equation}
   \int_0^1g(t)e^{i\nu' f(t)}dt,
   \end{equation}
where $f(t)$ and $g(t)$ are real functions:
$f(t)=\rho\ln t-\rho\ln(1-t)-\ln(1-\xi t)$ in both integrals. 
For large values of $\nu'$, the integrands are strongly oscillating functions
of $t$.
The function $f(t)$ has a ``saddle-point'' at
$t=t_0\equiv(1-\rho)$/2, where 
the first and second derivatives of $f$ are zero ($f'(t_0)=f''(t_0)=0$). 
For large values of $\nu'$, most of the integral value comes from the
contribution in a small region around $t_0$. 
This allows to consider only the leading terms in the Taylor's expansion
of $f$ and $g$ around $t_0$; that is,  $f(t)\simeq f(t_0)+\frac{1}{6}
f'''(t_0)(t-t_0)^3$, and $g(t)\simeq g(t_0)+g'(t_0)(t-t_0)$,
so that the integration interval can be extended from $-\infty$ to
$\infty$.
Using Eqs. (5.9.6) and (5.9.7) of Ref. \onlinecite{nist}, these integrals
become
{\small
   \begin{equation}
   I_A\simeq\left(\frac{1-\rho}{1+\rho}\right)^{1+i\nu'(1+\rho)}
   \left(\frac{(1-\rho^2)^2}{16\rho\nu'}\right)^{1/3}\frac{2\pi}
   {3^{2/3}\Gamma(\frac{2}{3})},
   \end{equation}
   \begin{equation}
   I_B\simeq-\frac{4i}{(1-\rho)^2}
   \left(\frac{1-\rho}{1+\rho}\right)^{1+i\nu'(1+\rho)}
   \left(\frac{(1-\rho^2)^2}{16\rho\nu'}\right)^{2/3}
   3^{1/6}\Gamma({\scriptstyle\frac{2}{3}}).
   \end{equation}
}
Using these values in Eq.~(\ref{eq015}), one obtains
   \begin{equation}\label{eq032}
   G\simeq\frac{1}{4\pi\sqrt{3}}.
   \end{equation}
This procedure can be justified only if the neglected terms in the expansions
are small in the region around the ``saddle-point''
where $\nu'(f(t)-f(t_0))\sim 1$. 
This condition can be written as $\nu\gg1/(1-\rho)$ and, in terms of 
normalized quatities, as $\hat\varepsilon^{3/2}\ll\hat \omega\ll 1$.
It is satisfied for region V in Fig. \ref{eps-omg-plane}.
%
%
\subsection{Summary of asymptotic expressions \label{summaryb}}
Summarizing the finding of Appendix B (Eqs. (\ref{eq026}), (\ref{eq008}),
and (\ref{eq032})), we have
\begin{widetext}
\begin{equation}
   G(\hat\varepsilon,\hat\omega)\simeq
   \frac{1}{2\pi}\frac{1/\sqrt{2\hat\varepsilon}}
   {1-e^{-2\pi/\sqrt{2\hat\varepsilon}}}
   \ln\frac{\sqrt{\hat\varepsilon+\hat\omega}+\sqrt{\hat\varepsilon}}
   {\sqrt{\hat\varepsilon+\hat\omega}-\sqrt{\hat\varepsilon}}
   \simeq\left\{\begin{array}{ll}
   {\displaystyle \frac{1}{\pi\sqrt{2\hat\omega}}} 
   & (\text{Region I}: \hat\omega\gg1\gg\hat\varepsilon)\\
   &\\
   {\displaystyle\frac{1}{2\pi^2}\sqrt{\frac{\hat\varepsilon}{\hat\omega}}}& 
   (\text{Region II}: \hat\omega\gg\hat\varepsilon\gg1) \\
   \end{array}\right.
\end{equation}
\begin{equation}
   G(\hat\varepsilon,\hat\omega)\simeq
   \frac{1}{4\pi^2}(\ln\frac{4\hat\varepsilon}{\hat\omega}
   -a(1/\sqrt{2\hat\varepsilon}))
   \simeq\left\{\begin{array}{ll}
      {\displaystyle\frac{1}{4\pi^2}\ln\frac{4\hat\varepsilon}{\hat\omega}}
      & (\text{Region III}: \hat\omega\ll\hat\varepsilon,1\ll\hat\varepsilon)\\
      &\\
      {\displaystyle\frac{1}{4\pi^2}\ln\frac{4\sqrt{2}\hat\varepsilon^{3/2}}
      {\hat\omega e^{\gamma}}}
      & (\text{Region IV}: \hat\omega\ll\hat\varepsilon^{3/2}\ll 1)\\
   \end{array}\right.
   \end{equation}
\begin{equation}
   G(\hat\varepsilon,\hat\omega)\simeq
   \frac{1}{4\pi\sqrt{3}}
   \hspace{6cm}
   (\text{Region V}: \hat\varepsilon^{3/2}\ll\hat\omega\ll 1)
\end{equation}
\end{widetext}

It is noteworthy that the asymptotic expression for region III, 
is also contained in Eq.~(\ref{eq026}),
despite the fact that  Eq.~(\ref{eq026}) can only be justified in
regions I and II. 
Therefore Eq.~(\ref{eq026}) applies to regions I, II, and III.
%
%
\section[C]{Numerical evaluation of integrals}
The integral in Eq.~(\ref{nu_final}) has the form
   \begin{equation}
   \int_0^{\infty}G\bigg(\frac{\varepsilon}{Z^2E_a},
   \frac{\hbar\omega}{Z^2E_a}\bigg)\left[
   \frac{1}{1+e^{\frac{\varepsilon-\mu}{kT}}}-
   \frac{1}{1+e^{\frac{\varepsilon+\hbar\omega-\mu}{kT}}}\right]d\varepsilon.
   \end{equation}
For fixed $\omega$ and $Z$, the function $G$ is finite for $\varepsilon=0$, and
grows logarithmically for large $\varepsilon$.  
The term in square brackets is maximum for
$\varepsilon=\varepsilon_{m}
\equiv\max(0,\mu-\frac12\hbar \omega)$, and decreases
exponentially ($\propto e^{-\varepsilon/kT}$) for large $\varepsilon$.
Consequently, the upper integration limit can be set to 
$\varepsilon_{m}+100\,kT$ with negligible numerical error.
The integration is carried out using an adaptive algorithm. 
The integration interval is decomposed into subintervals.
In each subinterval, the integral and the numerical error are evaluated 
simultaneously.
The subinterval with larger absolute error is halved into two smaller
subintervals.
This process is repeated until the global error estimate is below a 
given bound (in this work, a relative accuracy of $10^{-6}$).
For smooth real functions (like the ones in Eq.~(\ref{nu_final})), 
the subinterval quadrature is done using
the composite Simpson's rule with five points
(Eq.~(3.5.7) of Ref. \onlinecite{nist}),
and the numerical error is obtained as the difference between quadratures
with five and with three points.

A modified version of the above algorithm has be applied to evaluate
the complex integrals Eqs.~(\ref{eq038}) and (\ref{eq014}).
Notice that, in both integrals, the integrand is singular 
at $t=0$ and at $t=1$.
In the first subinterval $[0,h]$ ($h\ll1$),
the integrand can be approximated using a truncated Taylor series
   \begin{equation}\label{eq024}
   (1-t)^{-i\rho\nu'}(1-\xi t)^{-i\nu'-1}=
   1+g_1t+g_2t^2+g_3t^3+\dots+g_{N}t^{N},
   \end{equation}
and the integration is then carried out term by term. 
When this expression is applied to Eq.~(\ref{eq014}),
the first term in the series gives place to the improper integral
$\int_{0}^{h}t^{i\rho\nu'-1}dt$. 
In that case, as it was explained in Appendix A, we
have to take the limit: $\lim_{\varsigma\rightarrow0}
\int_{0}^{h}t^{i\rho\nu'+\varsigma-1}dt
=h^{i\rho\nu'}/({i\rho\nu'})$.
As an estimate of the error, one can use the contribution coming 
from the last term in Eq.~(\ref{eq024}).
The same procedure can be applied to the last subinterval $[1-h,1]$.
For moderate values of $\nu'$, Simpson's quadrature can be used for
the rest of subintervals.
However, for large values of $\nu'$, the integrand is a fast oscillatory 
function of $t$, and the use of the Simpson's quadrature would require
a prohibitive large number of function evaluations.
To overcome this limitation,
in each internal subinterval $[t_i,t_{i+1}]$, the complex 
integrand is approximated as
   \begin{equation}
   f(t)\simeq \bar f(u)\equiv(A+Bu+Cu^2)e^{i(D+E u)},
   \end{equation}
where $u$ is a local variable ranging from zero (at $t_i$) 
to one (at $t_{i+1}$), related to $t$ by
   \begin{equation}
   \frac{t-t_i}{t_{i+1}-t_i}=(1-F)u+Fu^2.
   \end{equation}
The six real constants $A$ to $F$ are determined by the
condition $\bar f=f$ at points
$t_i$, $\frac{1}{2}(t_i+t_{i+1})$, and $t_{i+1}$.
The integral across the subinterval can be evaluated in terms of 
elemental integrals
   \begin{equation}\label{eq017}
   \int_{t_i}^{t_{i+1}}f(t)dt\simeq 
   \int_{t_i}^{t_{i+1}}\bar f(u(t))dt=
   (t_{i+1}-t_i)e^{iD}$$$$\times\int_0^1(A+Bu+Cu^2)((1-F)+2Fu)e^{iEu}du.
   \end{equation}
Notice that, if $E\gg1$, a large number of oscillations occurs between
$t_i$ and $t_{i+1}$.
In each subinterval, the numerical error is evaluated as the difference 
between  direct application of Eq.~(\ref{eq017}) and the sum of values
obtained by dividing the subinterval in two halves.
%
%
\bibliography{article}

\begin{thebibliography}{32}%
\makeatletter
\providecommand \@ifxundefined [1]{%
 \@ifx{#1\undefined}
}%
\providecommand \@ifnum [1]{%
 \ifnum #1\expandafter \@firstoftwo
 \else \expandafter \@secondoftwo
 \fi
}%
\providecommand \@ifx [1]{%
 \ifx #1\expandafter \@firstoftwo
 \else \expandafter \@secondoftwo
 \fi
}%
\providecommand \natexlab [1]{#1}%
\providecommand \enquote  [1]{``#1''}%
\providecommand \bibnamefont  [1]{#1}%
\providecommand \bibfnamefont [1]{#1}%
\providecommand \citenamefont [1]{#1}%
\providecommand \href@noop [0]{\@secondoftwo}%
\providecommand \href [0]{\begingroup \@sanitize@url \@href}%
\providecommand \@href[1]{\@@startlink{#1}\@@href}%
\providecommand \@@href[1]{\endgroup#1\@@endlink}%
\providecommand \@sanitize@url [0]{\catcode `\\12\catcode `\$12\catcode
  `\&12\catcode `\#12\catcode `\^12\catcode `\_12\catcode `\%12\relax}%
\providecommand \@@startlink[1]{}%
\providecommand \@@endlink[0]{}%
\providecommand \url  [0]{\begingroup\@sanitize@url \@url }%
\providecommand \@url [1]{\endgroup\@href {#1}{\urlprefix }}%
\providecommand \urlprefix  [0]{URL }%
\providecommand \Eprint [0]{\href }%
\providecommand \doibase [0]{http://dx.doi.org/}%
\providecommand \selectlanguage [0]{\@gobble}%
\providecommand \bibinfo  [0]{\@secondoftwo}%
\providecommand \bibfield  [0]{\@secondoftwo}%
\providecommand \translation [1]{[#1]}%
\providecommand \BibitemOpen [0]{}%
\providecommand \bibitemStop [0]{}%
\providecommand \bibitemNoStop [0]{.\EOS\space}%
\providecommand \EOS [0]{\spacefactor3000\relax}%
\providecommand \BibitemShut  [1]{\csname bibitem#1\endcsname}%
\let\auto@bib@innerbib\@empty
\bibitem [{\citenamefont {Atzeni}\ and\ \citenamefont
  {Meyer{-ter}-Vehn}(2004)}]{Atzeni&MtV}%
  \BibitemOpen
  \bibfield  {author} {\bibinfo {author} {\bibfnamefont {S.}~\bibnamefont
  {Atzeni}}\ and\ \bibinfo {author} {\bibfnamefont {J.}~\bibnamefont
  {Meyer{-ter}-Vehn}},\ }\href@noop {} {\emph {\bibinfo {title} {The Physics of
  Inertial Fusion}}}\ (\bibinfo  {publisher} {Oxford Science Publications},\
  \bibinfo {address} {Oxford},\ \bibinfo {year} {2004})\BibitemShut {NoStop}%
\bibitem [{\citenamefont {Norreys}\ and\ \citenamefont
  {Drake}(2014)}]{Norreys}%
  \BibitemOpen
  \bibfield  {author} {\bibinfo {author} {\bibfnamefont {P.~A.}\ \bibnamefont
  {Norreys}}\ and\ \bibinfo {author} {\bibfnamefont {R.~P.}\ \bibnamefont
  {Drake}},\ }\href@noop {} {\bibfield  {journal} {\bibinfo  {journal} {New J.
  Phys.}\ }\textbf {\bibinfo {volume} {16}},\ \bibinfo {pages} {065007}
  (\bibinfo {year} {2014})}\BibitemShut {NoStop}%
\bibitem [{Sta()}]{Stanford_XFEL}%
  \BibitemOpen
  \href@noop {} {\emph {\bibinfo {title} {{Stanford XFEL:
  https://lcls.slac.stanford.edu/lcls-ii/science}}}}\BibitemShut {NoStop}%
\bibitem [{DES()}]{DESY_XFEL}%
  \BibitemOpen
  \href@noop {} {\emph {\bibinfo {title} {{DESY XFEL:
  https://www.xfel.eu/science}}}}\BibitemShut {NoStop}%
\bibitem [{\citenamefont {Hollebon}\ \emph {et~al.}(2018)\citenamefont
  {Hollebon}, \citenamefont {Ciricosta}, \citenamefont {Desjarlais},
  \citenamefont {Cacho}, \citenamefont {Spindloe}, \citenamefont {Springate},
  \citenamefont {Turcu}, \citenamefont {Warrk},\ and\ \citenamefont
  {Vinko}}]{Hollebon}%
  \BibitemOpen
  \bibfield  {author} {\bibinfo {author} {\bibfnamefont {P.}~\bibnamefont
  {Hollebon}}, \bibinfo {author} {\bibfnamefont {O.}~\bibnamefont {Ciricosta}},
  \bibinfo {author} {\bibfnamefont {M.}~\bibnamefont {Desjarlais}}, \bibinfo
  {author} {\bibfnamefont {C.}~\bibnamefont {Cacho}}, \bibinfo {author}
  {\bibfnamefont {C.}~\bibnamefont {Spindloe}}, \bibinfo {author}
  {\bibfnamefont {E.}~\bibnamefont {Springate}}, \bibinfo {author}
  {\bibfnamefont {I.}~\bibnamefont {Turcu}}, \bibinfo {author} {\bibfnamefont
  {J.}~\bibnamefont {Warrk}}, \ and\ \bibinfo {author} {\bibfnamefont
  {S.}~\bibnamefont {Vinko}},\ }\href@noop {} {\bibfield  {journal} {\bibinfo
  {journal} {to appear in Phys. Rev. E}\ } (\bibinfo {year}
  {2018})}\BibitemShut {NoStop}%
\bibitem [{\citenamefont {Sommerfeld}(1931)}]{Sommerfeld}%
  \BibitemOpen
  \bibfield  {author} {\bibinfo {author} {\bibfnamefont {A.}~\bibnamefont
  {Sommerfeld}},\ }\href@noop {} {\emph {\bibinfo {title} {Atombau und
  {S}pektrallinien}}}\ (\bibinfo  {publisher} {F. Vieweg {\&} Sohn},\ \bibinfo
  {address} {Braunschweig},\ \bibinfo {year} {1931})\BibitemShut {NoStop}%
\bibitem [{\citenamefont {Ashcroft}\ and\ \citenamefont
  {Mermin}(1976)}]{Ashcroft&Mermin}%
  \BibitemOpen
  \bibfield  {author} {\bibinfo {author} {\bibfnamefont {N.~W.}\ \bibnamefont
  {Ashcroft}}\ and\ \bibinfo {author} {\bibfnamefont {N.~D.}\ \bibnamefont
  {Mermin}},\ }\href@noop {} {\emph {\bibinfo {title} {Solid State Physics}}}\
  (\bibinfo  {publisher} {Saunders College Publishing},\ \bibinfo {address}
  {Ford Worth},\ \bibinfo {year} {1976})\BibitemShut {NoStop}%
\bibitem [{\citenamefont {Redmer}(1997)}]{Redmer}%
  \BibitemOpen
  \bibfield  {author} {\bibinfo {author} {\bibfnamefont {R.}~\bibnamefont
  {Redmer}},\ }\href@noop {} {\bibfield  {journal} {\bibinfo  {journal} {Phys.
  Rep.}\ }\textbf {\bibinfo {volume} {282}},\ \bibinfo {pages} {35} (\bibinfo
  {year} {1997})}\BibitemShut {NoStop}%
\bibitem [{\citenamefont {Ishikawa}\ \emph {et~al.}(1998)\citenamefont
  {Ishikawa}, \citenamefont {Felderhof}, \citenamefont {Blenski},\ and\
  \citenamefont {Cichocki}}]{Ishikawa1998}%
  \BibitemOpen
  \bibfield  {author} {\bibinfo {author} {\bibfnamefont {K.}~\bibnamefont
  {Ishikawa}}, \bibinfo {author} {\bibfnamefont {B.}~\bibnamefont {Felderhof}},
  \bibinfo {author} {\bibfnamefont {T.}~\bibnamefont {Blenski}}, \ and\
  \bibinfo {author} {\bibfnamefont {B.}~\bibnamefont {Cichocki}},\ }\href@noop
  {} {\bibfield  {journal} {\bibinfo  {journal} {J. Plasma Phys.}\ }\textbf
  {\bibinfo {volume} {60}},\ \bibinfo {pages} {787} (\bibinfo {year}
  {1998})}\BibitemShut {NoStop}%
\bibitem [{\citenamefont {Spitzer}(1962)}]{Spitzer}%
  \BibitemOpen
  \bibfield  {author} {\bibinfo {author} {\bibfnamefont {L.~J.}\ \bibnamefont
  {Spitzer}},\ }\href@noop {} {\emph {\bibinfo {title} {The physics of fully
  ionized plasmas}}}\ (\bibinfo  {publisher} {Wiley Interscience},\ \bibinfo
  {address} {New York},\ \bibinfo {year} {1962})\BibitemShut {NoStop}%
\bibitem [{\citenamefont {Kramers}(1923)}]{Kramers}%
  \BibitemOpen
  \bibfield  {author} {\bibinfo {author} {\bibfnamefont {H.~A.}\ \bibnamefont
  {Kramers}},\ }\href@noop {} {\bibfield  {journal} {\bibinfo  {journal}
  {London, Edinburgh, and Dublin Philosophical Magazine and Journal of
  Science}\ }\textbf {\bibinfo {volume} {49}},\ \bibinfo {pages} {836}
  (\bibinfo {year} {1923})}\BibitemShut {NoStop}%
\bibitem [{\citenamefont {Gaunt}(1930)}]{Gaunt1930}%
  \BibitemOpen
  \bibfield  {author} {\bibinfo {author} {\bibfnamefont {J.~A.}\ \bibnamefont
  {Gaunt}},\ }\href@noop {} {\bibfield  {journal} {\bibinfo  {journal} {Phil.
  Trans. R. Soc. London}\ }\textbf {\bibinfo {volume} {A229}},\ \bibinfo
  {pages} {163} (\bibinfo {year} {1930})}\BibitemShut {NoStop}%
\bibitem [{\citenamefont {Karzas}\ and\ \citenamefont {Latter}(1961)}]{KL1960}%
  \BibitemOpen
  \bibfield  {author} {\bibinfo {author} {\bibfnamefont {W.~J.}\ \bibnamefont
  {Karzas}}\ and\ \bibinfo {author} {\bibfnamefont {R.}~\bibnamefont
  {Latter}},\ }\href@noop {} {\bibfield  {journal} {\bibinfo  {journal}
  {Astrophys. J. Suppl.}\ }\textbf {\bibinfo {volume} {6}},\ \bibinfo {pages}
  {167} (\bibinfo {year} {1961})}\BibitemShut {NoStop}%
\bibitem [{\citenamefont {Sutherland}(1998)}]{Sutherland1998}%
  \BibitemOpen
  \bibfield  {author} {\bibinfo {author} {\bibfnamefont {R.~S.}\ \bibnamefont
  {Sutherland}},\ }\href@noop {} {\bibfield  {journal} {\bibinfo  {journal}
  {Mon. Not. R. Astron. Soc.}\ }\textbf {\bibinfo {volume} {300}},\ \bibinfo
  {pages} {321} (\bibinfo {year} {1998})}\BibitemShut {NoStop}%
\bibitem [{\citenamefont {van Hoof}\ \emph {et~al.}(2014)\citenamefont {van
  Hoof}, \citenamefont {Williams}, \citenamefont {Volk}, \citenamefont
  {Chatzikos}, \citenamefont {Ferland}, \citenamefont {Lykins}, \citenamefont
  {Porter},\ and\ \citenamefont {Wang}}]{Hoof2014}%
  \BibitemOpen
  \bibfield  {author} {\bibinfo {author} {\bibfnamefont {P.~A.~M.}\
  \bibnamefont {van Hoof}}, \bibinfo {author} {\bibfnamefont {R.~J.~R.}\
  \bibnamefont {Williams}}, \bibinfo {author} {\bibfnamefont {K.}~\bibnamefont
  {Volk}}, \bibinfo {author} {\bibfnamefont {M.}~\bibnamefont {Chatzikos}},
  \bibinfo {author} {\bibfnamefont {G.~J.}\ \bibnamefont {Ferland}}, \bibinfo
  {author} {\bibfnamefont {M.}~\bibnamefont {Lykins}}, \bibinfo {author}
  {\bibfnamefont {R.~L.}\ \bibnamefont {Porter}}, \ and\ \bibinfo {author}
  {\bibfnamefont {Y.}~\bibnamefont {Wang}},\ }\href@noop {} {\bibfield
  {journal} {\bibinfo  {journal} {Mon. Not. R. Astron. Soc.}\ }\textbf
  {\bibinfo {volume} {444}},\ \bibinfo {pages} {420} (\bibinfo {year}
  {2014})}\BibitemShut {NoStop}%
\bibitem [{\citenamefont {de~Avillez}\ and\ \citenamefont
  {Breitschwerdt}(2015)}]{Avillez2015}%
  \BibitemOpen
  \bibfield  {author} {\bibinfo {author} {\bibfnamefont {M.}~\bibnamefont
  {de~Avillez}}\ and\ \bibinfo {author} {\bibfnamefont {D.}~\bibnamefont
  {Breitschwerdt}},\ }\href@noop {} {\bibfield  {journal} {\bibinfo  {journal}
  {Astron. Astrophys.}\ }\textbf {\bibinfo {volume} {580}},\ \bibinfo {pages}
  {A124} (\bibinfo {year} {2015})}\BibitemShut {NoStop}%
\bibitem [{\citenamefont {Reif}(2009)}]{Reif}%
  \BibitemOpen
  \bibfield  {author} {\bibinfo {author} {\bibfnamefont {F.}~\bibnamefont
  {Reif}},\ }\href@noop {} {\emph {\bibinfo {title} {Fundamentals of
  statistical and thermal physics}}}\ (\bibinfo  {publisher} {Waveland Press
  Inc.},\ \bibinfo {address} {Long Grove, Illinois},\ \bibinfo {year}
  {2009})\BibitemShut {NoStop}%
\bibitem [{\citenamefont {Einstein}(1916)}]{Einstein}%
  \BibitemOpen
  \bibfield  {author} {\bibinfo {author} {\bibfnamefont {A.}~\bibnamefont
  {Einstein}},\ }\href@noop {} {\bibfield  {journal} {\bibinfo  {journal}
  {Deutsche Physikalische Gesellschaft}\ }\textbf {\bibinfo {volume} {18}},\
  \bibinfo {pages} {318} (\bibinfo {year} {1916})}\BibitemShut {NoStop}%
\bibitem [{\citenamefont {{Abramowitz (editor)}}\ and\ \citenamefont {{Stegun
  (editor)}}(1964)}]{abramovitz}%
  \BibitemOpen
  \bibfield  {author} {\bibinfo {author} {\bibfnamefont {M.}~\bibnamefont
  {{Abramowitz (editor)}}}\ and\ \bibinfo {author} {\bibfnamefont {I.~A.}\
  \bibnamefont {{Stegun (editor)}}},\ }\href@noop {} {\emph {\bibinfo {title}
  {Handbook of mathematical functions with formulas, graphs, and mathematical
  tables}}}\ (\bibinfo  {publisher} {U. S. Government Printing Office},\
  \bibinfo {address} {Washington, D. C},\ \bibinfo {year} {1964})\BibitemShut
  {NoStop}%
\bibitem [{\citenamefont {{Olver (editor)}}\ \emph {et~al.}(2010)\citenamefont
  {{Olver (editor)}}, \citenamefont {{Lozier (editor)}}, \citenamefont
  {{Boisvert (editor)}},\ and\ \citenamefont {{Clark (editor)}}}]{nist}%
  \BibitemOpen
  \bibfield  {author} {\bibinfo {author} {\bibfnamefont {F.~W.~J.}\
  \bibnamefont {{Olver (editor)}}}, \bibinfo {author} {\bibfnamefont {D.~W.}\
  \bibnamefont {{Lozier (editor)}}}, \bibinfo {author} {\bibfnamefont {R.~F.}\
  \bibnamefont {{Boisvert (editor)}}}, \ and\ \bibinfo {author} {\bibfnamefont
  {C.~W.}\ \bibnamefont {{Clark (editor)}}},\ }\href@noop {} {\emph {\bibinfo
  {title} {{NIST} Handbook of Mathematical Functions}}}\ (\bibinfo  {publisher}
  {Cambridge University Press and National Institute of Standards and
  Technology},\ \bibinfo {year} {2010})\BibitemShut {NoStop}%
\bibitem [{\citenamefont {Berestetskii}, \citenamefont {Lifshitz},\ and\
  \citenamefont {Pitaevski}(1971)}]{Landau&Lifshitz}%
  \BibitemOpen
  \bibfield  {author} {\bibinfo {author} {\bibfnamefont {V.}~\bibnamefont
  {Berestetskii}}, \bibinfo {author} {\bibfnamefont {E.}~\bibnamefont
  {Lifshitz}}, \ and\ \bibinfo {author} {\bibfnamefont {L.~P.}\ \bibnamefont
  {Pitaevski}},\ }\href@noop {} {\emph {\bibinfo {title} {Relativistic Quantum
  Theory}}},\ \bibinfo {series} {Course in Theoretical Physics}, Vol.~\bibinfo
  {volume} {4}\ (\bibinfo  {publisher} {Pergamon Press},\ \bibinfo {address}
  {Oxford},\ \bibinfo {year} {1971})\BibitemShut {NoStop}%
\bibitem [{\citenamefont {Krainov}(2000)}]{KrainovI}%
  \BibitemOpen
  \bibfield  {author} {\bibinfo {author} {\bibfnamefont {V.~P.}\ \bibnamefont
  {Krainov}},\ }\href@noop {} {\bibfield  {journal} {\bibinfo  {journal} {J.
  Phys. B: At. Mol. Opt. Phys.}\ }\textbf {\bibinfo {volume} {33}},\ \bibinfo
  {pages} {1585} (\bibinfo {year} {2000})}\BibitemShut {NoStop}%
\bibitem [{\citenamefont {Krainov}(2001)}]{KrainovII}%
  \BibitemOpen
  \bibfield  {author} {\bibinfo {author} {\bibfnamefont {V.~P.}\ \bibnamefont
  {Krainov}},\ }\href@noop {} {\bibfield  {journal} {\bibinfo  {journal} {J.
  Exp. Theor. Physics}\ }\textbf {\bibinfo {volume} {92}},\ \bibinfo {pages}
  {960} (\bibinfo {year} {2001})}\BibitemShut {NoStop}%
\bibitem [{\citenamefont {Chandrasekhar}(1967)}]{Chandra}%
  \BibitemOpen
  \bibfield  {author} {\bibinfo {author} {\bibfnamefont {S.}~\bibnamefont
  {Chandrasekhar}},\ }\href@noop {} {\emph {\bibinfo {title} {An Introduction
  to the Study of Stellar Structure}}}\ (\bibinfo  {publisher} {Dover
  Publications Inc.},\ \bibinfo {address} {New York},\ \bibinfo {year}
  {1967})\BibitemShut {NoStop}%
\bibitem [{\citenamefont {Christensen-Dalsgaard}\ \emph
  {et~al.}(1996)\citenamefont {Christensen-Dalsgaard}, \citenamefont {Däppen},
  \citenamefont {Ajukov}, \citenamefont {Anderson}, \citenamefont {Antia},
  \citenamefont {Basu}, \citenamefont {Baturin}, \citenamefont {Berthomieu},
  \citenamefont {Chaboyer}, \citenamefont {Chitre}, \citenamefont {Cox},
  \citenamefont {Demarque}, \citenamefont {Donatowicz}, \citenamefont
  {Dziembowski}, \citenamefont {Gabriel}, \citenamefont {Gough}, \citenamefont
  {Guenther}, \citenamefont {Guzik}, \citenamefont {Harvey}, \citenamefont
  {Hill}, \citenamefont {Houdek}, \citenamefont {Iglesias}, \citenamefont
  {Kosovichev}, \citenamefont {Leibacher}, \citenamefont {Morel}, \citenamefont
  {Proffitt}, \citenamefont {Provost}, \citenamefont {Reiter}, \citenamefont
  {{Rhodes Jr.}}, \citenamefont {Rogers}, \citenamefont {Roxburgh},
  \citenamefont {Thompson},\ and\ \citenamefont {Ulrich}}]{Christensen}%
  \BibitemOpen
  \bibfield  {author} {\bibinfo {author} {\bibfnamefont {J.}~\bibnamefont
  {Christensen-Dalsgaard}}, \bibinfo {author} {\bibfnamefont {W.}~\bibnamefont
  {Däppen}}, \bibinfo {author} {\bibfnamefont {S.~V.}\ \bibnamefont {Ajukov}},
  \bibinfo {author} {\bibfnamefont {E.~R.}\ \bibnamefont {Anderson}}, \bibinfo
  {author} {\bibfnamefont {H.~M.}\ \bibnamefont {Antia}}, \bibinfo {author}
  {\bibfnamefont {S.}~\bibnamefont {Basu}}, \bibinfo {author} {\bibfnamefont
  {V.~A.}\ \bibnamefont {Baturin}}, \bibinfo {author} {\bibfnamefont
  {G.}~\bibnamefont {Berthomieu}}, \bibinfo {author} {\bibfnamefont
  {B.}~\bibnamefont {Chaboyer}}, \bibinfo {author} {\bibfnamefont {S.~M.}\
  \bibnamefont {Chitre}}, \bibinfo {author} {\bibfnamefont {A.~N.}\
  \bibnamefont {Cox}}, \bibinfo {author} {\bibfnamefont {P.}~\bibnamefont
  {Demarque}}, \bibinfo {author} {\bibfnamefont {J.}~\bibnamefont
  {Donatowicz}}, \bibinfo {author} {\bibfnamefont {W.~A.}\ \bibnamefont
  {Dziembowski}}, \bibinfo {author} {\bibfnamefont {M.}~\bibnamefont
  {Gabriel}}, \bibinfo {author} {\bibfnamefont {D.~O.}\ \bibnamefont {Gough}},
  \bibinfo {author} {\bibfnamefont {D.~B.}\ \bibnamefont {Guenther}}, \bibinfo
  {author} {\bibfnamefont {J.~A.}\ \bibnamefont {Guzik}}, \bibinfo {author}
  {\bibfnamefont {J.~W.}\ \bibnamefont {Harvey}}, \bibinfo {author}
  {\bibfnamefont {F.}~\bibnamefont {Hill}}, \bibinfo {author} {\bibfnamefont
  {G.}~\bibnamefont {Houdek}}, \bibinfo {author} {\bibfnamefont {C.~A.}\
  \bibnamefont {Iglesias}}, \bibinfo {author} {\bibfnamefont {A.~G.}\
  \bibnamefont {Kosovichev}}, \bibinfo {author} {\bibfnamefont {J.~W.}\
  \bibnamefont {Leibacher}}, \bibinfo {author} {\bibfnamefont {P.}~\bibnamefont
  {Morel}}, \bibinfo {author} {\bibfnamefont {C.~R.}\ \bibnamefont {Proffitt}},
  \bibinfo {author} {\bibfnamefont {J.}~\bibnamefont {Provost}}, \bibinfo
  {author} {\bibfnamefont {J.}~\bibnamefont {Reiter}}, \bibinfo {author}
  {\bibfnamefont {E.~J.}\ \bibnamefont {{Rhodes Jr.}}}, \bibinfo {author}
  {\bibfnamefont {F.~J.}\ \bibnamefont {Rogers}}, \bibinfo {author}
  {\bibfnamefont {I.~W.}\ \bibnamefont {Roxburgh}}, \bibinfo {author}
  {\bibfnamefont {M.~J.}\ \bibnamefont {Thompson}}, \ and\ \bibinfo {author}
  {\bibfnamefont {R.~K.}\ \bibnamefont {Ulrich}},\ }\href@noop {} {\bibfield
  {journal} {\bibinfo  {journal} {Science}\ }\textbf {\bibinfo {volume}
  {272}},\ \bibinfo {pages} {1286} (\bibinfo {year} {1996})}\BibitemShut
  {NoStop}%
\bibitem [{\citenamefont {Ichimaru}(1994)}]{Ichimaru}%
  \BibitemOpen
  \bibfield  {author} {\bibinfo {author} {\bibfnamefont {S.}~\bibnamefont
  {Ichimaru}},\ }\href@noop {} {\emph {\bibinfo {title} {Statistical Plasma
  Physics}}},\ Vol.\ \bibinfo {volume} {II: Condensed Plasmas}\ (\bibinfo
  {publisher} {Addison Wesley},\ \bibinfo {address} {Reading, MA},\ \bibinfo
  {year} {1994})\BibitemShut {NoStop}%
\bibitem [{\citenamefont {Wesson}(1987)}]{Wesson}%
  \BibitemOpen
  \bibfield  {author} {\bibinfo {author} {\bibfnamefont {J.}~\bibnamefont
  {Wesson}},\ }\href@noop {} {\emph {\bibinfo {title} {Tokamaks}}}\ (\bibinfo
  {publisher} {Clarendon Press},\ \bibinfo {address} {Oxford},\ \bibinfo {year}
  {1987})\ p.~\bibinfo {pages} {40}\BibitemShut {NoStop}%
\bibitem [{Note1()}]{Note1}%
  \BibitemOpen
  \bibinfo {note} {One may conjecture that this limit corresponds to the limit
  $\omega \rightarrow 0$, had we treated Debye screening correctly, and one is
  tempted to substitute $\protect \sqrt {\omega ^2+\omega _p^2}$ for $\omega $
  under the logarithm in Eq.~{\ref {nu_eff regionIII}}.}\BibitemShut {Stop}%
\bibitem [{\citenamefont {Pearson}(2009)}]{pearson}%
  \BibitemOpen
  \bibfield  {author} {\bibinfo {author} {\bibfnamefont {J.}~\bibnamefont
  {Pearson}},\ }\emph {\bibinfo {title} {Computation of Hypergeometric
  Functions}},\ \href@noop {} {Ph.D. thesis},\ \bibinfo  {school} {University
  of Oxford} (\bibinfo {year} {2009})\BibitemShut {NoStop}%
\bibitem [{\citenamefont {Brussaard}\ and\ \citenamefont {{H. C. van de
  Hulst}}(1962)}]{BH1962}%
  \BibitemOpen
  \bibfield  {author} {\bibinfo {author} {\bibfnamefont {P.~J.}\ \bibnamefont
  {Brussaard}}\ and\ \bibinfo {author} {\bibnamefont {{H. C. van de Hulst}}},\
  }\href@noop {} {\bibfield  {journal} {\bibinfo  {journal} {Rev. Mod.
  Physics}\ }\textbf {\bibinfo {volume} {34}},\ \bibinfo {pages} {507}
  (\bibinfo {year} {1962})}\BibitemShut {NoStop}%
\bibitem [{\citenamefont {Press}\ \emph {et~al.}(1992)\citenamefont {Press},
  \citenamefont {Teukolsky}, \citenamefont {Vetterling},\ and\ \citenamefont
  {Flannery}}]{recipes}%
  \BibitemOpen
  \bibfield  {author} {\bibinfo {author} {\bibfnamefont {W.~H.}\ \bibnamefont
  {Press}}, \bibinfo {author} {\bibfnamefont {S.~A.}\ \bibnamefont
  {Teukolsky}}, \bibinfo {author} {\bibfnamefont {W.~T.}\ \bibnamefont
  {Vetterling}}, \ and\ \bibinfo {author} {\bibfnamefont {B.~P.}\ \bibnamefont
  {Flannery}},\ }\href@noop {} {\emph {\bibinfo {title} {Numerical Recipes in
  C}}}\ (\bibinfo  {publisher} {Cambridge University Press},\ \bibinfo
  {address} {Cambridge},\ \bibinfo {year} {1992})\BibitemShut {NoStop}%
\bibitem [{\citenamefont {Shkarofsky}, \citenamefont {Johnston},\ and\
  \citenamefont {Bachynski}(1966)}]{shkarofsky}%
  \BibitemOpen
  \bibfield  {author} {\bibinfo {author} {\bibfnamefont {I.~P.}\ \bibnamefont
  {Shkarofsky}}, \bibinfo {author} {\bibfnamefont {T.~W.}\ \bibnamefont
  {Johnston}}, \ and\ \bibinfo {author} {\bibfnamefont {M.~P.}\ \bibnamefont
  {Bachynski}},\ }\href@noop {} {\emph {\bibinfo {title} {The particle kinetics
  of plasmas}}}\ (\bibinfo  {publisher} {Addison-Wesley},\ \bibinfo {address}
  {Reading, Massachusetts},\ \bibinfo {year} {1966})\BibitemShut {NoStop}%
\end{thebibliography}%
\end{document}